\documentclass[preprint]{aastex}

\newcommand{\CaII}{\ion{Ca}{2}}
\newcommand{\Ks}{\ensuremath{K_{S}}}
\newcommand{\JKs}{\ensuremath{J\!-\!K_{S}}}
\newcommand{\JH}{\ensuremath{J\!-\!H}}
\newcommand{\HKs}{\ensuremath{H\!-\!K_{S}}}
\newcommand{\VKs}{\ensuremath{V\!-\!K_{S}}}
\newcommand{\VJ}{\ensuremath{V\!-\!J}}
\newcommand{\by}{\ensuremath{b\!-\!y}}
\newcommand{\BV}{{\it BV}}
\newcommand{\uvby}{{\it uvby}}
\newcommand{\etal}{et~al.}
\newcommand{\Msun}{\ensuremath{\mathrm{M}_{\sun}}}
\newcommand{\Mv}{\ensuremath{M_{V}}}
\newcommand{\MvsdB}{\ensuremath{M_{V\!\mathrm{,sdB}}}}
\newcommand{\hour}{\ensuremath{^{\mathrm{h}}}}
\newcommand{\minute}{\ensuremath{^{\mathrm{m}}}}
\newcommand{\PopI}{Pop~\scshape{i}}
\newcommand{\PopII}{Pop~\scshape{ii}}
\newcommand{\chisq}{\ensuremath{\chi^{2}}}
\newcommand{\chisqr}{\ensuremath{\chi^{2}_{\mathrm{R}}}}

\shorttitle{Hot Subdwarf Stars in 2MASS}
\shortauthors{Stark {\&} Wade}

\begin{document}

\title{Single and Composite Hot Subdwarf Stars in the Light of 2MASS
Photometry}

\author{M.~A.~Stark and Richard A.~Wade}
\affil{Department of Astronomy and Astrophysics, The Pennsylvania
 State University, University Park, PA 16802}
\email{stark@astro.psu.edu, wade@astro.psu.edu}

\begin{abstract}
Utilizing the Two Micron All Sky Survey (2MASS) Second Incremental
Data Release Catalog, we have retrieved near-IR magnitudes for several
hundred hot subdwarfs (sdO and sdB stars) drawn from the {\it
Catalogue of Spectroscopically Identified Hot Subdwarfs}
{\citep*[][1992]{Kilkenny}}.  This sample size greatly exceeds that of
previous studies of hot subdwarfs.  Examining 2MASS photometry alone
or in combination with visual photometry (Johnson {\BV} or
Str\"{o}mgren {\uvby}) available in the literature, we show that it is
possible to identify hot subdwarf stars that exhibit atypically red IR
colors that can be attributed to the presence of an unresolved late
type companion.  Utilizing this large sample, we attempt for the first
time to define an approximately volume limited sample of hot
subdwarfs.  We discuss the considerations, biases, and difficulties in
defining such a sample.

We find that, of the hot subdwarfs in Kilkenny {\etal}, about 40{\%}
in a magnitude limited sample have colors that are consistent with the
presence of an unresolved late type companion.  Binary stars are
over-represented in a magnitude limited sample.  In an approximately
volume limited sample the fraction of composite-color binaries is
about 30{\%}.
\end{abstract}



\section{Introduction} \label{sec:Intro}

Hot subdwarf stars as a class have effective temperatures exceeding
$\sim$20,000~K, with surface gravities higher and luminosities lower
than main sequence stars of the same temperature.  Spectroscopically,
hot subdwarfs are divided into two major subtypes: subdwarf~B (sdB)
and subdwarf~O (sdO).  SdB stars have hydrogen-dominated atmospheres
with $T_{\mathrm{eff}}\approx 24,000$--30,000~K and exhibit a
relatively small dispersion in luminosity
{\citep{Heber,Saffer,deBoer}}.  They often exhibit weak lines of
Helium, Carbon, and other atomic species in comparison with
Population~{\scshape{i}} main sequence stars.  SdO stars have
helium-enriched atmospheres with $T_{\mathrm{eff}} > 30,000$~K and
exhibit a wider range of $\log g$ and luminosity than the sdB stars
{\citep{Heber,Saffer,deBoer}}.  Hot subdwarf stars (and sdB stars in
particular) are the primary source of the ``UV upturn'' (or ``UVX'')
seen in elliptical galaxies and spiral galaxy bulges
{\citep[e.g.,][]{Greggio,Brown97,Brown00,OConnell}}.

Based on their temperatures and magnitudes, {\citet{HumasonZwicky}}
proposed that the sdB stars are the field equivalent of {\citet{Baade}}
population~{\scshape{ii}} B stars (now known as ``extended
horizontal branch\footnote{The term ``extended horizontal branch'' was
coined by {\citet{G+S}}.},'' or ``EHB,'' stars) which are seen
in some globular cluster HR diagrams at the blue end of the horizontal
branch.  As such, sdB stars in the field are understood to be core
helium burning objects with very low hydrogen envelope masses
(M$_{\mathrm{env}} \lesssim 0.02$~\Msun) and total masses of
$\mathrm{M}\approx0.5-0.55$~{\Msun} {\citep{Caloi,Saffer}}.  Some sdO
stars have colors and luminosities that are consistent with the
hottest part of the EHB (least massive H envelopes).  However, there
are a number of potential evolutionary paths, e.g.~post-AGB evolution,
that lead through the rather broad region of the HR diagram occupied
by sdO stars.  All hot subdwarfs, regardless of their prior evolution,
are thought to be direct progenitors of white dwarfs, although they
constitute only a small fraction of all stars that evolve to become
white dwarfs.

While the current evolutionary status of sdB stars is relatively well
understood, how they actually arrived at the EHB with such low
envelope masses is not understood.  The spectroscopically inferred
masses of field sdB stars have a very small dispersion.  If they
evolved directly from red giant progenitors, then their initial masses
are constrained to be $\lesssim 2$~{\Msun}, leading to a requirement
that they lose up to $\sim$1.5~{\Msun} on the red giant branch before
or right at core helium ignition {\citep{Saffer}}.  One scenario for
such extreme mass loss suggests that sdB stars lost significant
envelope mass due to binary interactions on the red giant branch
{\citep*{Mengel}}.  If this were the sole channel for forming sdB
stars, {\it all} sdB stars would be expected to be found in {\it
close} binary systems --- that is systems where the current separation
is of the same order as the radius of the sdB progenitor star near the
tip of the red giant branch, or R~$\approx100$~R$_{\sun}\approx0.5$~AU
{\citep{Maxted}}.  Companions to sdB stars have been reported by
numerous surveys, involving a variety of photometric and spectroscopic
techniques, although these often have been of limited scope.  These
include: searches for flux excesses at long wavelengths, or flat
energy distributions {\citep*[e.g.,][]{AWLBF,Allard,TUM,UllaThejll}};
radial velocity variations {\citep*[e.g.,][]{Maxted,SGB}}; the
presence of a G-band, {\CaII} absorption, or other spectral features
indicative of late type stars
{\citep*[e.g.,][]{Ferguson,Bixler,Theissen,JeffPoll,SGB,Maxted}}; and
even direct high-resolution imaging
{\citep*[e.g.,][]{TTJ,HSTimaging}}.

Some of these survey reports have included an attempt to deduce the
fraction of sdBs that are in binaries.  The results have ranged from
50 to 100{\%}, depending on differing specific assumptions made about
the nature and physical parameters of the sdB stars and/or their
companions.  For example, the evolutionary status of the companion
stars is typically not known.  Some authors have preferred main
sequence companions and less luminous sdB stars
{\citep[i.e.,][]{Allard,Ferguson,Bixler}}, whereas others have
preferred slightly evolved, sub-giant or giant companions and more
luminous sdB stars {\citep[i.e.,][]{JeffPoll,Theissen}}.  The
correction for ``unseen'' fainter companions (such as dM stars) in
determining the total binary fraction is greatly affected by this
choice. Thus, if the observed (G--K type) companions in composite
systems are subgiants, the number of fainter unobserved companions
(extrapolated using an assumed initial mass function) must be very
large, and the resulting binary fraction is close to 100{\%}.  On the
other hand, if the G--K companions are main sequence stars, the number
of fainter unobserved companions is much smaller (because the lower
inferred luminosity for the hot subdwarf is less effective at masking
the light of such companions). The result is a binary fraction as low
as 50{\%}.  However, only the radial velocity or imaging studies give
information about whether the sdBs that do have companions are in {\it
close} binary systems.

The systematic survey of the entire sky that was carried out by the
Two Micron All Sky Survey\footnote{http://www.ipac.caltech.edu/2mass/}
(2MASS) provides the opportunity to further explore the question of
whether all sdB stars (and sdO stars, to the extent that they
represent the EHB) are in binary systems. This can be done by
combining cataloged photometry at visible wavelengths with the 2MASS
near-infrared imaging photometry. In addition to comparing visible and
near-IR magnitudes, near-infrared colors for hot subdwarfs (e.g.,
{\JKs}) can be used on their own as a second indicator of a binary
system.  In this paper we collect readily available visible and
near-IR flux measurements of hot subdwarfs and identify those whose
colors indicate the presence of a late type companion. We thus
determine the fraction of subdwarfs that exist in composite spectrum
binaries. (Note that not all classes of companion star will be
identified this way, in particular white dwarf companions.) We extract
visible photometry from an existing catalog of hot subdwarf stars, and
we use IR photometry obtained from the 2MASS Second Incremental Data
Release (2IDR)
Catalog\footnote{http://www.ipac.caltech.edu/2mass/releases/second/index.html}.
We look for IR colors that are ``redder'' than what would be expected
for a normal hot subdwarf and which can be attributed to the presence
of an unresolved late type companion. This method is similar to that
used by {\citet{TUM}} and {\citet{UllaThejll}}, but our sample is much
larger and the IR photometry is more precise.

In {\S\ref{sec:Sample}} we describe our sample of hot subdwarfs
including: improved coordinates, retrieval of 2MASS infrared
photometry, availability and sources of visual photometry, and adopted
classifications for objects in the initial sample.  In
{\S\ref{sec:HotSdColor}} we examine and model a magnitude limited
sample of hot subdwarfs using two color diagrams and the distribution
in {\JKs} color.  In {\S\ref{sec:Vol}} we introduce an
approximately ``volume limited'' sample of hot subdwarfs and compare
it to the magnitude limited sample.  Finally, in {\S\ref{sec:Summary}}
we summarize our findings.

\section{The Sample} \label{sec:Sample}

For this investigation we studied hot subdwarf stars listed in the
{\it Catalogue of Spectroscopically Identified Hot Subdwarfs}
{\citep*{Kilkenny}} as updated and expanded in an electronic version
by Kilkenny, c.\,1992, kindly made available to us by D.~Kilkenny.
This catalog (hereinafter referred to as KHD\nocite{Kilkenny})
contains 1527 entries, of which 302 entries were added since the 1988
publication.  Hot subdwarfs from the PG survey of UV-excess objects
{\citep*{PG}} are an important component of the catalog (1017 objects
with PG designations, either discovered or recovered by that survey).
More recent or ongoing surveys, including the Hamburg Schmidt (HS) and
Edinburgh-Cape (EC) surveys are not included.  In the catalog
sub-sample of 593 objects that is the subject of this paper, about
84{\%} of the stars are to be found in the 1988 printed version of
KHD\nocite{Kilkenny}, and about half of the sub-sample stars added to
KHD\nocite{Kilkenny} between the 1988 and 1992 versions are from the
Carnegie Southern Observatory H\&K survey {\citep{Beers}}.  About
71{\%} of the entries in KHD\nocite{Kilkenny} have northern
Declinations, the same percentage as in the sub-sample studied
here. The sub-sample is thus broadly representative of the parent
catalog, for stars bright enough to be included in the 2MASS point
source database.  A number of important selection effects bias the
parent catalog's contents, however, since it is compiled from many
different sources each with its own criteria for inclusion or
discovery (see also {\S\ref{sec:JKs}}), but KHD\nocite{Kilkenny} is
the most complete single compilation of field hot subdwarfs available,
especially for the northern sky and for relatively bright objects.

While the KHD\nocite{Kilkenny} catalog contains all varieties of hot
subdwarfs, we are primarily interested in the more numerous sdB stars.
As mentioned above, sdB stars are understood to be relatively
homogeneous and probably have a common evolution history from the zero
age extended horizontal branch (ZAEHB) and beyond, while sdO stars
likely follow multiple evolutionary pathways and might be expected to
be less homogeneous and to have less simply explained properties.  In
the following sections, while we focus on the sdB stars, we will take
the sdO stars ``along for the ride'' and comment on the two groups
either separately (sdB or sdO), or combined as ``hot subdwarfs'' in
general (both sdB and sdO).  We further discuss the adopted
classifications of objects from KHD\nocite{Kilkenny} and our treatment
of them in {\S\ref{sec:class}}.

\subsection{Coordinates} \label{sec:Coord}

To make a comparison of the KHD\nocite{Kilkenny} data with 2MASS,
accurate coordinates on a consistent system are required.  For each
entry we verified the object's position by referring whenever possible
to original published finding charts or by contacting knowledgeable
observers, then locating the object on a chart prepared from the USNO
A2.0 Catalog\footnote{http://www.nofs.navy.mil/data/fchpix/}
{\citep{usno}}.  We adopted coordinates from USNO A2.0, except for
bright objects missing from that source, in which case we used
coordinates from the Tycho-2 {\citep{T2}} or Hipparcos
{\citep{Perryman,Tycho}} catalogs.  These updated ICRS 2000.0
coordinates are accurate to a few arcseconds ($\sim2''$ or better) at
the epoch of observation; in many cases this represents a very
substantial improvement over the coordinates reported by
KHD\nocite{Kilkenny}.  During the time in which we collected the 2MASS
data, we had confirmed and improved coordinates for 1459 objects (not
counting 11 duplicate entries), while for 57 objects we had not yet
positively identified the correct star.  These remaining 57
unrecovered objects are not included in the present study.  The full
collection of improved coordinates will be published elsewhere (Wade
{\&} Stark, in preparation).

\subsection{Infrared Photometry} \label{sec:2MASS}

Infrared photometry ($J$, $H$, and \Ks) from the 2MASS 2IDR (which
covers $\sim$45{\%} of the sky) was collected by performing searches
within $\sim10''$ of the the updated coordinates.  Visual verification
was performed for crowded fields by comparing images from the
Digitized Sky Survey\footnote{http://archive.stsci.edu/dss/index.html}
(DSS) with images from 2MASS (most hot subdwarfs are uncrowded, with
no other comparably bright field stars within $\sim30''$).  Objects
are included in 2MASS if they have photometric uncertainty $\leq
0.155$~mag (signal-to-noise ratio, S/N$\geq$7) in at least one of the
three bands ($J$, $H$, or \Ks).  This results in reliable photometry
for point sources brighter than $J = 15.8$, $H = 15.1$, and $\Ks =
14.3$\footnote{However, according to the {\it Explanatory Supplement
to the 2MASS Second Incremental Data Release} (located at:
http://www.ipac.caltech.edu/2mass/releases/second/doc/explsup.html),
at high Galactic latitude, the 2MASS Point Source Catalog contains
accurate detections $0.5\!-\!1.0$ magnitudes fainter than these
limits.}. {\JKs} errors for our objects are typically less than
$\sim$0.2~mag, with an average {\JKs} error of $\sim$0.12~mag for our
sample.  Of the initial sample (1459 objects), we were able to
retrieve 2MASS photometry for {571} hot subdwarfs (593 before removing
22 objects classified as non-subdwarfs).  Of the {571} hot subdwarfs,
only {385} had both $J$ and {\Ks} magnitudes listed.  Of the other
{186} objects, {180} have measurements of $J$ but only 95{\%} confidence upper
limits on {\Ks}, and {six} have detections of a nearby companion to the
subdwarf (but not the subdwarf itself).  Also in the observations
returned from 2MASS, a small fraction (few percent) of subdwarfs have
two possible identifications within $\sim10\!-\!20''$ of the input
coordinates.  The hot subdwarf members of these visual doubles were
included in the present study since both finding charts and 2MASS
images were available to identify the correct member of the pair.
These visual doubles and their interpretation as possible physical
binaries will be discussed in more detail in future work.

Table~\ref{allcoord} lists the identifiers and coordinates for those
objects from KHD\nocite{Kilkenny} that were found in 2MASS.  Columns 1
and 2 list a sequence number\footnote{The sequence number, assigned by
us, is not intended to be used as a new designator in the literature,
but is useful in navigating the KHD\nocite{Kilkenny} catalog and our
Tables~\ref{allcoord} and {\ref{all}}.} and the first object
identifier given in KHD\nocite{Kilkenny}.  Columns 3 and 4 contain the
J2000 coordinates based on the USNO A2.0 system.  Column 5 is the
adopted classification for the hot subdwarf (see {\S\ref{sec:class}}).
Column 6 contains notes on individual objects as well as cross
references for duplicate entries (see Table~\ref{allcoord} footnotes).
Column 7 lists an additional identifier for the hot subdwarf, if
needed, using nomenclature consistent with
SIMBAD\footnote{http://vizier.u-strasbg.fr/, operated at CDS,
Strasbourg, France}.  For example, the designator CSHK~22876$-35$ used
by KHD\nocite{Kilkenny} for star {\#}8 in this table is equivalent to
the SIMBAD designator BPS~CS~22876$-35$, and the designator
KPD~00330$+5229$ used by KHD\nocite{Kilkenny} for star {\#}55 in this
table is equivalent to the SIMBAD designator KPD~0033$+5229$.
PG~23595$+1942$ used by KHD\nocite{Kilkenny} for star {\#}1527 in this
table is equivalent to the SIMBAD designator PG~2359$+197$.  From now
on, when we refer to a specific object from the sample, we will use
both its sequence number (for ease in cross-referencing our
Tables~\ref{allcoord} and {\ref{all}} with each other and with
KHD\nocite{Kilkenny}) and a SIMBAD-compliant designator.  If no other
ID is given in the last column, the identifier from
KHD\nocite{Kilkenny} (column 2) is consistent with SIMBAD.  In this
way, one SIMBAD-compliant designator for each star is available for
cross-reference.  The exceptions are the ``HS'' star {\#}390
HS~1000$+4704$ {\citep*{HJW}}, the ``SC'' star {\#}1165 SC~1721$-3338$
from {\citet{Reid}}, and the ``LBQS'' stars from
{\citet[][{\#}41, 565, and 1405]{Berg}} which are
not presently recognized by SIMBAD; we note that SIMBAD {\it does} use
the LBQS designator for quasars from {\citet*{HFC}}.

\subsection{Visual Photometry} \label{sec:visual}

Visual photometry ({\BV} or {\uvby}) for the hot subdwarfs was taken
from KHD\nocite{Kilkenny}, {\citet{Allard}}, {\citet{Landolt}},
{\citet{Mooney}}, {\citet*{WMG}}, and/or the Tycho-1 Catalog
{\citep{T1}}.  For the photometry taken from KHD\nocite{Kilkenny},
when more than one set of photoelectric measurements was given, we
favored Johnson {\BV} over Str\"{o}mgren {\uvby}, unless the only
{\BV} measurement was reported as ``uncertain'' (then {\uvby} was
used).  If only a single ``uncertain'' {\BV} measurement was available
for an object (and no Str\"{o}mgren data were found), then that
uncertain measurement was nevertheless used and treated as exact.  For
consistency, we took the first of these measurements listed in
KHD\nocite{Kilkenny}, unless it was uncertain, and even if additional
measurements differed significantly from the first.  Those hot
subdwarfs with only photographic visual photometry were not included
in the optical-infrared color space examination of hot subdwarfs
discussed in {\S\ref{sec:ColorColor}--\ref{sec:Model}} (however, those
with 2MASS 2IDR detections are included in the objects discussed in
{\S\ref{sec:JKs}}).  For stars listed in {\citet{Allard}},
{\citet{Landolt}}, or {\citet{WMG}} the magnitudes from those sources
were used in place of any reported in KHD\nocite{Kilkenny}.
Str\"{o}mgren {\uvby} was taken from {\citet{Mooney}} for two objects:
{\#}320 PG~09146$+0004$ and {\#}1327 PG~21325$+1235$ in preference to
uncertain $V$ and no {\BV} (respectively) in KHD\nocite{Kilkenny}.
Tycho-1 {\BV} magnitudes {\citep{T1,Tycho}} were taken for {\#}234
BD~$+34\arcdeg1543$, which had only a single $V$ magnitude listed in
KHD\nocite{Kilkenny}.  No further literature search was undertaken to
locate additional visual photometry.  No errors are reported for the
photometry given in KHD\nocite{Kilkenny} and we did not refer to
original references to locate errors for each individual observation.
However typical errors are $\lesssim0.02$~mag for those surveys comprising
the majority of the visual photometry
{\citep[i.e.,][]{Allard,PG,Mooney,Wesemael,WMG}}--- thus, for
simplicity, all measurements of visual photometry were treated as
exact.

When only Str\"{o}mgren photometry was available (or was taken
in place of uncertain {\BV}), we assumed that $y=V$, and {\by} was
converted to {\bv} using:
\[(\bv) =1.584 \times (\by) + 0.681 \times m_{1} - 0.116 \]
where: $m_{1} \equiv (v\!-\!b) - (\by)$~\citep{Turner}.  Two objects
have only synthetic $b$ and $y$ reported: {\#}1076 Balloon 083500007 
and {\#}1095 Balloon 083600004.  For these two objects, {\bv} was 
estimated by setting $m_1 = 0$ in this equation.

Tycho-1 $B$ and $V$ magnitudes for one object, {\#}234 BD~$+34\arcdeg1543$,
were converted to Johnson $B$ and $V$ magnitudes using:
\begin{eqnarray*}
 V_{J}     & = & V_{T} - 0.090 \times (B_{T}\!-\!V_{T}) \\
 (\bv)_{J} & = & 0.850 \times (B_{T}\!-\!V_{T})
\end{eqnarray*}
This conversion is valid over $-0.2 < (B_{T}\!-\!V_{T}) < 1.8$, and
generally gives errors $<0.015$ in $V_{J}$ and $<0.05$ in $({\bv})_{J}$
{\citep{Tycho}}.

Of the {571} reported hot subdwarfs found in the 2MASS 2IDR, we were
able to collect photoelectric {\it V} or {\it y} magnitudes for {348}
stars, and {\bv} or the {\by} equivalent for {339} of these {348}
stars.

All collected IR and visual photometry is presented in
Table~\ref{all}.  Columns 1 and 2 list the sequence number and first
identifier given in KHD\nocite{Kilkenny}.  Columns 3--8 contain the
2MASS magnitudes ($J$, $H$, and {\Ks}) and their respective 1$\sigma$
errors.  If no error is listed, then the star was not detected in that
band, and the quoted magnitude is a 95{\%} confidence upper limit.
Columns 9--10 are the calculated {\JKs} and combined 1$\sigma$ error
for it. If no error is listed then the reported {\JKs} value is the
maximum color index when an upper limit on {\Ks} is reported, and it
is the minimum color index for an upper limit on $J$.  Column 11 lists
the $V$, $y$, or photographic visual magnitude of the hot subdwarf.
Column 12 contains the {\bv} color index for the hot subdwarf (derived
from Str\"{o}mgren {\uvby} or Tycho-1 {\BV} in necessary, as discussed
above).  Column 13 is the adopted classification for the hot subdwarf
(see {\S\ref{sec:class}}).  Column 14 contains additional notes, which
are explained in the footnotes to Table~\ref{all}.

\subsection{Classifications} \label{sec:class}

During our analysis, we divided the total sample of hot subdwarfs with
2MASS 2IDR data into two sub-samples: ``sdB/sd'' and ``sdO/sdOB''
based on their classifications in KHD\nocite{Kilkenny}.  For
consistency, we took the first classification given, even if there
were additional conflicting classifications.  Our ``sdB/sd'' group
comprises those objects classified as sdB, sdB--O, and sd.  Our
``sdO/sdOB'' group comprises sdO, sdO(A), sdO(B), sdO(C), sdO(D), and
sdOB.  There is ongoing discussion in the community regarding
appropriate multi-dimensional classification schemes for hot subdwarf
stars, which likely will depend on the spectral resolution used
{\citep{Jeffery,Drilling00,Drilling03}}.  Meanwhile, we have made no
attempt to improve upon, regularize, or rationalize the various
classifications summarized in KHD\nocite{Kilkenny}.

As noted earlier, 22 additional objects were excluded from our sample
because they are not subdwarfs. These objects are classified: HBB
(three~objects; Tables~\ref{allcoord} and~\ref{all} sequence~{\#}:
350, 642, and 662), DA (two; {\#}265 and 588), non-sd (nine; {\#}269,
287, 574, 676, 734, 794, 1138, 1443, and 1501), only as
``bin?''~(three; {\#}732, 1076, and 1095), or had no classification
(four; {\#}398, 702, 884, and 967) in KHD\nocite{Kilkenny}. Lastly,
{\#}276 KUV~08388+4029 is listed as ``sdB'' in KHD\nocite{Kilkenny},
however, SIMBAD lists this object as a Seyfert 1 galaxy at
$z\approx0.15$.  We verified its AGN nature with a low-resolution
spectrum obtained from the Hobby-Eberly Telescope\footnote{The
Hobby-Eberly Telescope is operated by McDonald Observatory on behalf
of The University of Texas at Austin, the Pennsylvania State
University, Stanford University, Ludwig-Maximilians-Universit\"{a}t
M\"{u}nchen, and Georg-August-Universit\"{a}t G\"{o}ttingen. The
Marcario Low Resolution Spectrograph is a joint project of the
Hobby-Eberly Telescope partnership and the Instituto de
Astronom\'{\i}a de la Universidad Nacional Autonoma de M\'{e}xico.},
which clearly showed the H$\alpha$ and H$\beta$ broad emission lines
at wavelengths consistent with a redshift $z\approx0.15$.

Hereafter, when we refer to ``sdB stars'' we mean our ``sdB/sd''
group, when we refer to ``sdO stars'' we mean our ``sdO/sdOB'' group,
and when we refer to ``hot subdwarfs'' we mean the entire sample (sdB
and sdO) excluding the non-subdwarfs just mentioned.  A summary of the
numbers and types of subdwarfs at each stage in the data gathering
process is presented in Table~\ref{Breakdown}.

\section{Hot Subdwarfs in Color Space} \label{sec:HotSdColor}

\subsection{Optical-IR Color-Color Plots} \label{sec:ColorColor}

After examining various combinations of color indices, we chose to
focus on (\JKs) vs.~(\bv) and (\JKs) vs.~(\VKs).  These two
combinations allow us to extract most of the information from the
available photometry.  Substituting {\JH} in place of {\JKs} shows the
same features and trends. Likewise, substituting {\VJ} for {\VKs}
shows the same features and trends.  Using {\HKs} did not provide any
additional information.

The locations of single and composite hot subdwarfs in color space
were determined by selecting samples of hot subdwarfs reported as
single and composite from the literature.  A total of 69 reportedly
single stars and 37 reported composites that had 2MASS detections were
taken from the literature
{\citep[KHD\nocite{Kilkenny};][]{Ferguson,Beers,Allard,TUM,UllaThejll}}
or a list of hot subdwarfs spectroscopically observed and kindly
provided to us by E.~M.~Green.  After removing from this list stars
with incomplete photometry and stars classified as ``non-sd'' by
KHD\nocite{Kilkenny}, 33 (24~sdB, 9~sdO) of the reported composites
and 59 (48~sdB, 11~sdO) of the reported singles remained and are used
here to determine the composite and single loci (these stars are
flagged in Table~\ref{all} with a ``$s$'' or ``$c$'' in the last
column).

\begin{figure}
  \epsscale{0.77}
  \rotatebox{-90}{\plotone{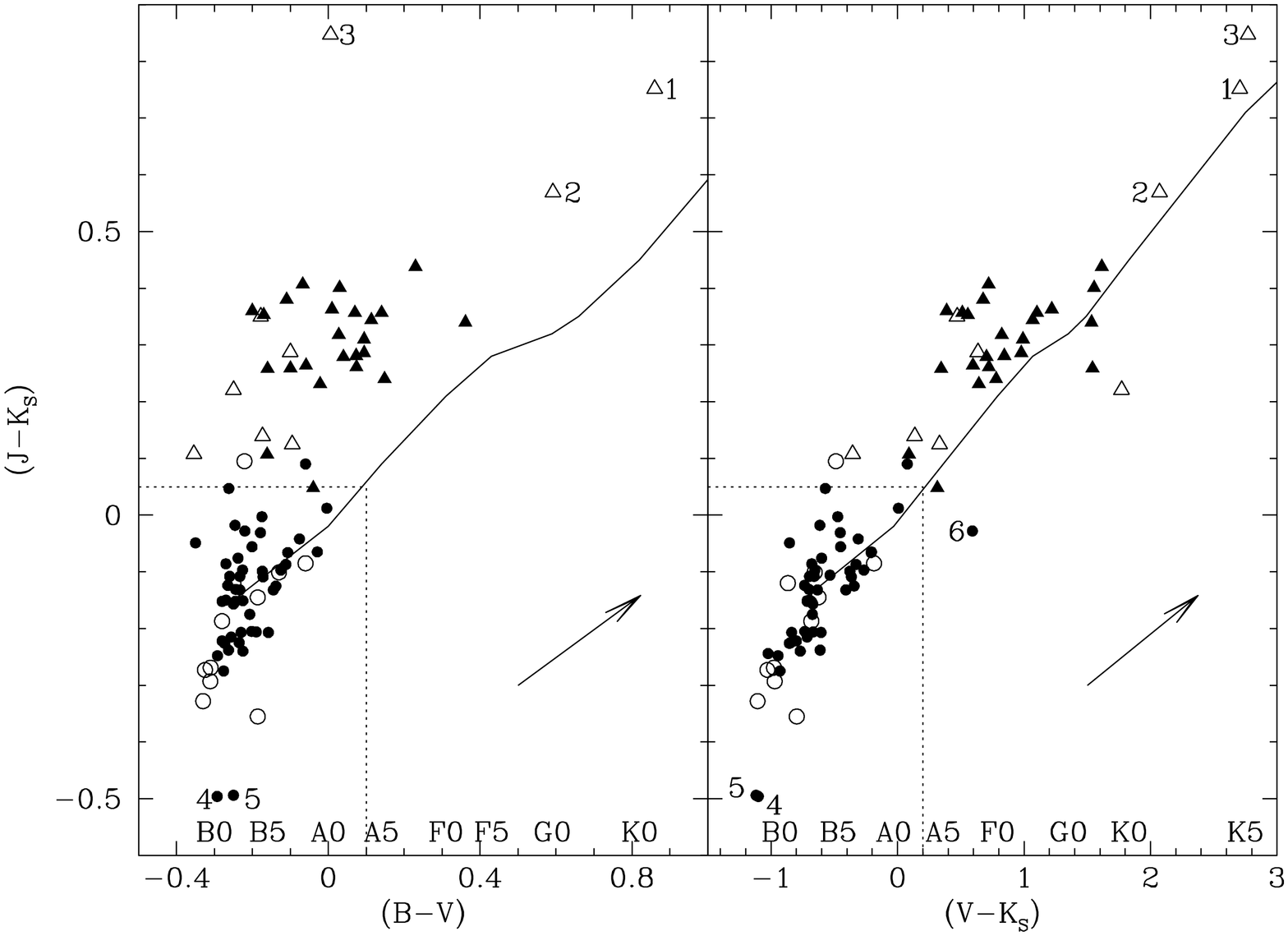}}

  \caption{Color-color diagrams for reported single (circles) and
  composite (triangles) hot subdwarfs.  Filled points represent hot
  subdwarfs in the sdB group (48 single, 24 composite), open points
  represent stars in the sdO group (10 single, 8 composite). The left
  panel shows (\JKs) vs.~(\bv), while the right panel shows
  (\JKs) vs.~(\VKs).  The diagonal solid line in each panel
  indicates the location of the Population~{\scshape{i}} main sequence
  (with the approximate {\PopI} spectral types indicated along the
  {\bv} and {\VKs} axes).  The single hot subdwarfs are mostly
  contained within the box defined by: $\bv \lesssim +0.1$,
  $\VKs \lesssim +0.2$, and $\JKs \lesssim +0.05$, which is shown in
  each panel with a dotted line.  Objects labeled 1--5 in both panels,
  and 6 in the right panel, are discussed further in the text
  (\S\ref{sec:ColorColor}).  The arrows indicate color shift due to
  one magnitude of extinction at $V$.  \label{Known} }
\end{figure}

When plotted in (\JKs, \bv) or (\JKs, \VKs) color spaces relative to
the location of the {\PopI} main sequence (Figure~\ref{Known}), the
reported single hot subdwarfs (both sdB and sdO) occupy the same area
in color space as other O and B stars, with the sdOs in general having
slightly bluer colors than the sdBs.  Most of the single hot subdwarfs
occupy the region: $(\bv) \lesssim+0.1$, $(\VKs) \lesssim+0.2$, and
$(\JKs) \lesssim+0.05$. This region is marked by a box on
Figure~\ref{Known}.  On the other hand, reported composite hot
subdwarfs (both sdB and sdO) occupy a separate and distinct region of
color space; they have redder colors than single subdwarf stars.
Composite subdwarfs show a consistent and quite large color difference
from the ``typical'' single subdwarf in {\JKs}. The difference in
{\VKs} is also large in the mean, but the distributions of single and
composite systems overlap slightly.  The least average difference is
in {\bv} (Figure~\ref{Known}).

These patterns can be understood if the hot subdwarf's companion star
in the composite systems is cool and relatively faint at visible
wavelengths.  The bluest color index, {\bv}, will be relatively
unaffected by the additional light from the companion.  The
visual--infrared index, {\VKs}, will betray a large infrared
``excess'' compared to a single hot subdwarf, but there will be a
large spread because the index depends on the competing light from the
two stars at $V$.  The pure infrared index, {\JKs}, will be dominated
by the cool companion, whose color is very different from that of the
hot star.

Three reported composite sdOs have distinctly different colors in
Figure~\ref{Known} compared with all of the other reported composites.
The first (labeled ``1'' in Figure~\ref{Known}) is {\#}1355
{\hbox{BD~$-3\arcdeg5357$}} (or FF~Aqr, with $\bv=+0.895$,
$\JKs=+0.752$, and $\VKs=+2.838$), an eclipsing binary (orbital period
$\sim$9.2~days) classified sdOB+G8III {\citep{Dworetsky,Etzel}}.  The
second (``2'') is {\#}1457 {\hbox{BD~$-7\arcdeg5977$}} (with
$\bv=+0.541$, $\JKs=+0.569$, and $\VKs=+2.129$), classified
sdOB+K0IV--III by {\citet{Viton}}.  The third (``3'') is {\#}127
{\hbox{PG~0205$+134$}} (with $\bv=+0.006$, $\JKs=+0.847$, and
$\VKs=+2.771$), classified ``sdOB comp'' by {\citet{PG}}, and further
refined to sdOB+M3.5(IV?) by {\citet{Allard}} and {\citet{WMG}}.  All
of these cool companions are atypically luminous for sdO-type
composites, which otherwise have main sequence or subgiant companions
reported, so it is not unreasonable for these three objects to have
exceptionally red colors.

The two sdB stars labeled ``4'' ({\#}822 PG~1439$-013$) and ``5''
({\#}744 TON~183) in Figure~\ref{Known}, have anomalously blue {\JKs}
colors, $\JKs =-0.496\pm0.148$ and $-0.494\pm0.090$ respectively
(1$\sigma$ errors).  These colors could be due to photometric
errors,  or they could be intrinsic to the objects themselves.  
The color index {\JH} generally shows the same trend as {\JKs}, and these
two stars have essentially the same {\JH} colors as all other
single sdBs ($\JH =-0.097$ and $-0.196$).  Possibly, these unusual
{\JKs} colors are due to photometric errors in {\Ks}.

One reported single sdB star, {\#}483 TON~64 (labeled ``6'' in the
right plot of Figure~\ref{Known}), has a rather red value for {\VKs}
(+0.592), yet appears ``normal'' in the other colors ($\bv =-0.220$,
$\JKs =-0.028$).  Its unusual position may be due to an error in the
standardization of the $V$ magnitude.  The two measurements for the
visual magnitude of this star in KHD\nocite{Kilkenny} are discrepant:
$V=15.86$ and $y=14.59$.  Our adopted procedure is to favor $V$ over
$y$; if however we were to use $y=14.59$ in place of $V=15.86$, we
would have $\VKs=-0.678$ (again assuming $y\!-\!\Ks=\VKs$), which
places this star near the center of the single subdwarf locus.
Nevertheless for consistency, we continue to use the value of $V$ for
this object.

\begin{figure}
  \epsscale{0.77}
  \rotatebox{-90}{\plotone{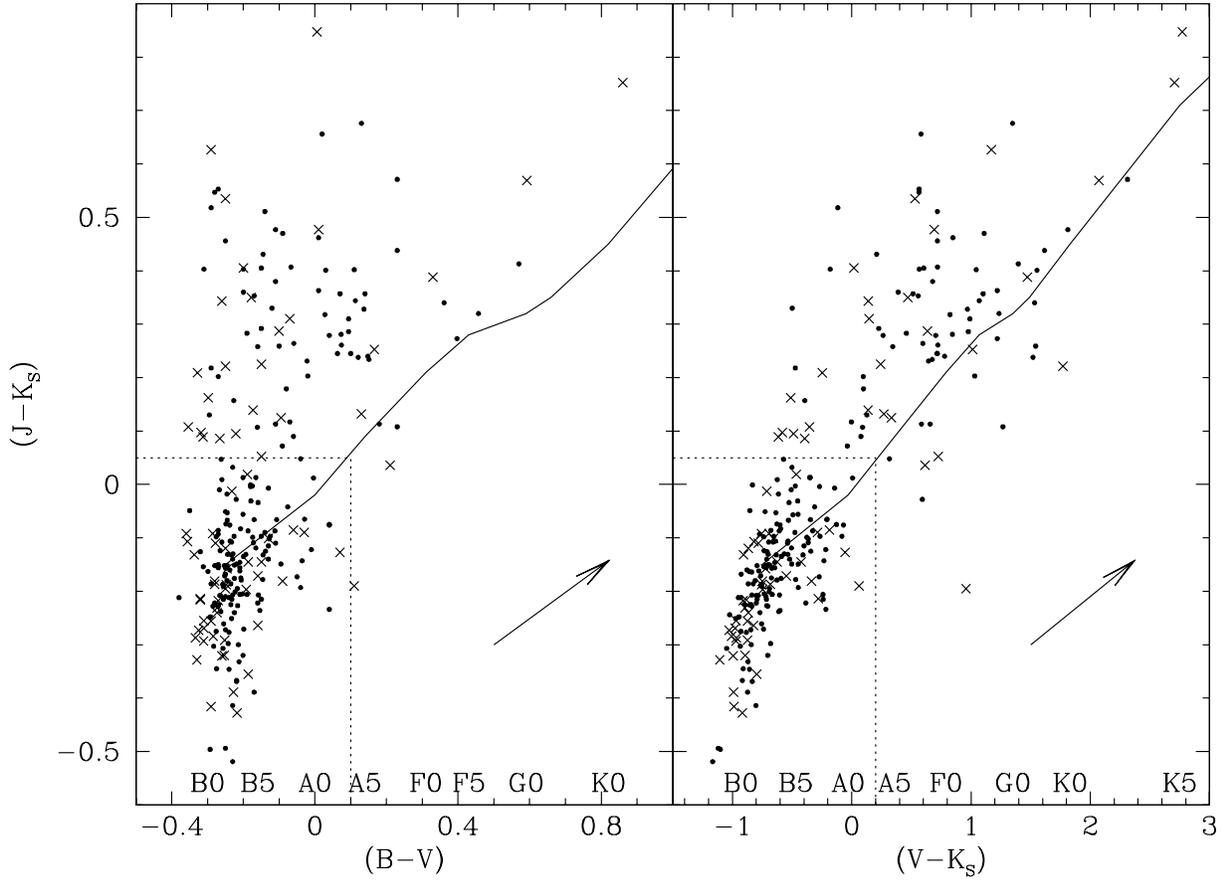}}

  \caption{Same as Figure~\ref{Known}, except all hot subdwarfs with
  2MASS 2IDR and visual photometry are shown.  SdBs are denoted with
  circles and sdOs with crosses (``$\times$'').  There are 201 sdBs
  and 71 sdOs in the left panel and 206 sdBs and 71 sdOs in the right;
  these totals include the reported composite and single stars of
  Figure~\ref{Known}.  One sdB, {\#}462 TON~56, has a published $\bv
  =-0.90$ which places it off the left side of the left panel, this
  {\BV} photometry from {\citet{PG}} is listed as uncertain; however,
  it is shown in the right panel ($\VKs =+0.263$, $\JKs =+0.279$).
  Four other sdBs are missing from the left panel (but not the right),
  because they have no $B$ (or equivalent) measurements reported.
  \label{allsdB} }
\end{figure}

Now placing all of the 277 hot subdwarfs with $V$ (273 with {\bv}) and
$J$ and {\Ks} measurements in color space (Figure~\ref{allsdB}), we
observe that they occupy the same areas in color space as defined by
the smaller sample of known composites and singles. (Objects with only
upper limits on {\Ks} from 2MASS are not plotted in
Figure~\ref{allsdB}.)  There is some additional scatter, which is most
likely due to errors in the photometry.  This scatter could, in
principle, also be due in part to differences in interstellar
extinction and reddening from object to object.  Additionally, some
hot subdwarfs are known to pulsate such as the V361~Hya
stars\footnote{also known unofficially as ``EC~14026'' stars}
{\citep{ODonoghue}} and the ``PG~1716'' stars {\citep{LongPeriod}}.
In the case of the V361~Hya stars, the pulsation amplitudes are
generally less than 0.01~mag and typically only a few milli-mags with
typical periods of 100--250~seconds
{\citep[i.e.,][]{KKOS,Ostensen,LongPeriod}}.  The ``PG~1716'' stars
have peak-to-peak amplitudes $\lesssim0.05$ magnitudes over timescales
of about an hour {\citep{LongPeriod}}.  Light variations due to both
of these pulsations are typically less than other sources of error.

Figure~\ref{allsdB} suggests that candidates for composite systems
(hot subdwarf+late type star) can be identified and targeted for
future study using {\bv} or {\VKs} in combination with
{\JKs}. Unfortunately, in order to use this technique, accurate visual
photometry is needed in addition to IR photometry.

\subsection{Modeling the Distribution in Optical-IR Color-Color
 Space} \label{sec:Model}

To test the interpretation offered above, that the colors of single
and composite hot subdwarfs may be understood if the companions are
cool stars, we attempted to reproduce the location of composite
subdwarfs in color space by combining the colors of a typical sdB with
those of late type {\PopI} main sequence stars.  Color indices for
late type stars were taken from {\citet{Johnson}}; for this exercise
we assumed $K \approx \Ks$.  The ``typical'' colors for sdBs were
taken from the center of the clump of reported single stars in
Figure~\ref{Known}.  We used $(\bv)_{\mathrm{sdB}}\approx-0.25$,
$(\VKs)_{\mathrm{sdB}}\approx-0.75$,
$(\JKs)_{\mathrm{sdB}}\approx-0.15$. Based on the spread seen in
Figure~\ref{Known}, however, it is likely that single sdB stars in
fact have spreads in color index on the order of
$\Delta(\bv)\sim\pm0.1$, $\Delta(\VKs)\sim\pm0.2$, and
$\Delta(\JKs)\sim\pm0.15$, as might be expected from variations in the
physical properties of the sdBs (most notably the effective
temperature); some of the spread in {\VKs} and {\JKs} is however due
to limited IR photometric precision.

The colors were combined, varying the fraction of the light at $V$
that arises from the companion.  The resultant locus passes among the
known composite sdBs with contributions from the companion of
$\sim$10--60{\%} at $V$, and spectral types G--K
(Figure~\ref{colortrack}).  These fractional contributions and
companion types agree with those reported by {\citet{Allard}} for
sdBs, and {\citet{Ferguson}} and {\citet*{Orosz}} for hot sdBs and
sdOs.  While this general analysis also fits the sdOs, the unreddened
colors of a ``typical'' sdO should be somewhat bluer than those of a
sdB, since sdOs are hotter than sdBs.

The loci are plotted with the entire subdwarf sample in
Figure~\ref{Allcolortrack}.

\begin{figure}
  \epsscale{0.77}
  \rotatebox{-90}{\plotone{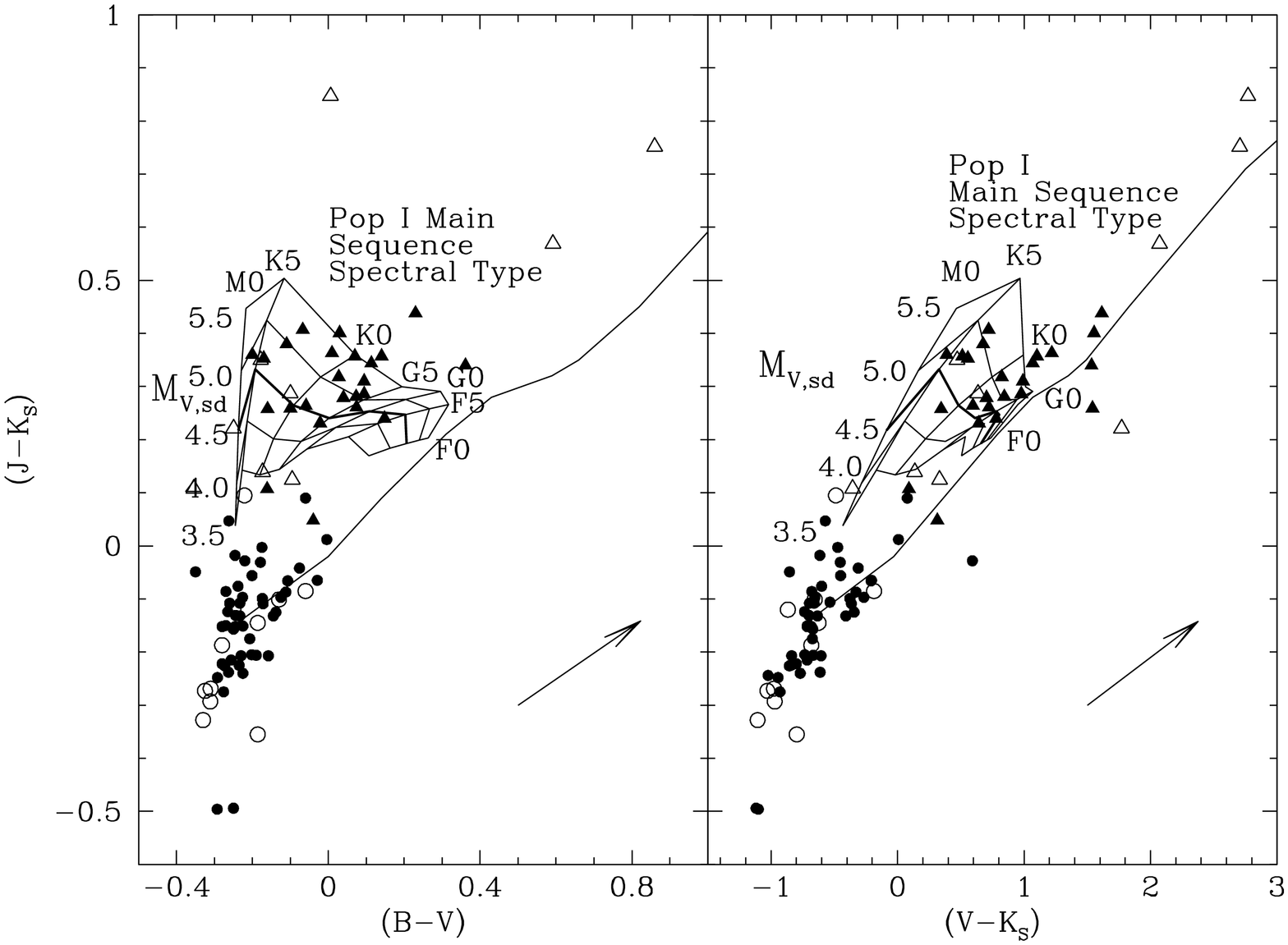}}

  \caption{Grid of composite colors computed by combining the light
  from a hot subdwarf ($\bv \approx -0.25$, $\VKs \approx -0.75$,
  $\JKs \approx -0.15$) with that of ``typical'' {\PopI} main
  sequence stars {\citep[F0, F5, G0, G5, K0, K5, and M0;][]{Johnson}}
  and assuming values of {\Mv} for the subdwarf ({\MvsdB}~=~3.5, 4.0,
  4.5, 5.0, and 5.5) --- the line for {\MvsdB}~=~4.5 has been darkened
  to help guide the eye.  The lines for G5 and F5 have been left off
  the right plot for clarity.  Circles are the reported single
  subdwarfs (solid: sdB, open: sdO) and triangles are the reported
  composite subdwarfs (solid: sdB, open: sdO) from Figure~\ref{Known}.
  The large arrows indicate the color change due to one magnitude of
  reddening. \label{MScolortrack} }
\end{figure}

By introducing assumed absolute magnitudes along with colors, we can
synthesize composite colors specifically for sdB+late type main
sequence stars.  We took colors and {\Mv} for {\PopI} main sequence
stars from {\citet{Johnson}} and {\citet{Cox}} and assumed colors and
{\MvsdB} for the sdB component. The sdB colors used in the previous
calculation were again adopted.  The absolute magnitudes of sdB stars
are poorly constrained, with reported values spanning the range from
$\MvsdB \approx3.5\!-\!6.2$
{\citep{Allard,Bixler,Ferguson,Liebert,Moehler,Orosz,Thejll}}.  We
therefore treated {\MvsdB} as a parameter, and varied it (choosing
$\MvsdB=3.5$, 4.0, 4.5, 5.0, and 5.5) to see its effect on the
location of the synthesized composites in color space
(Figure~\ref{MScolortrack}).  We found that the locus of synthetic
composites passes among the observed composite systems and best
characterizes the observed companions as having {\PopI} spectral types
G--K when $\MvsdB \approx4.5\!-\!5.0$~mag (if the companions lie on
the main sequence).  It should be noted, however, that the same
pattern of color loci can be created by fixing {\MvsdB} and instead
varying the absolute magnitudes of the companions from their main
sequence values.  If G--K type companions were subgiants, they would
be more luminous by perhaps $\sim$2 magnitudes, and would therefore
give the same pattern with a more luminous sdB, $\MvsdB
\approx2.5-3.0$~mag (but we would not expect to see evolved M--type
companions).  In nature, there is likely to be some range in absolute
magnitude for both the sdB stars and their companions. (In addition to
the variations of luminosity and bolometric correction along the
zero-age EHB, sdB stars should evolve to higher luminosity on
timescales of $10^8$~years.)

\begin{figure}
  \epsscale{0.77}
  \rotatebox{-90}{\plotone{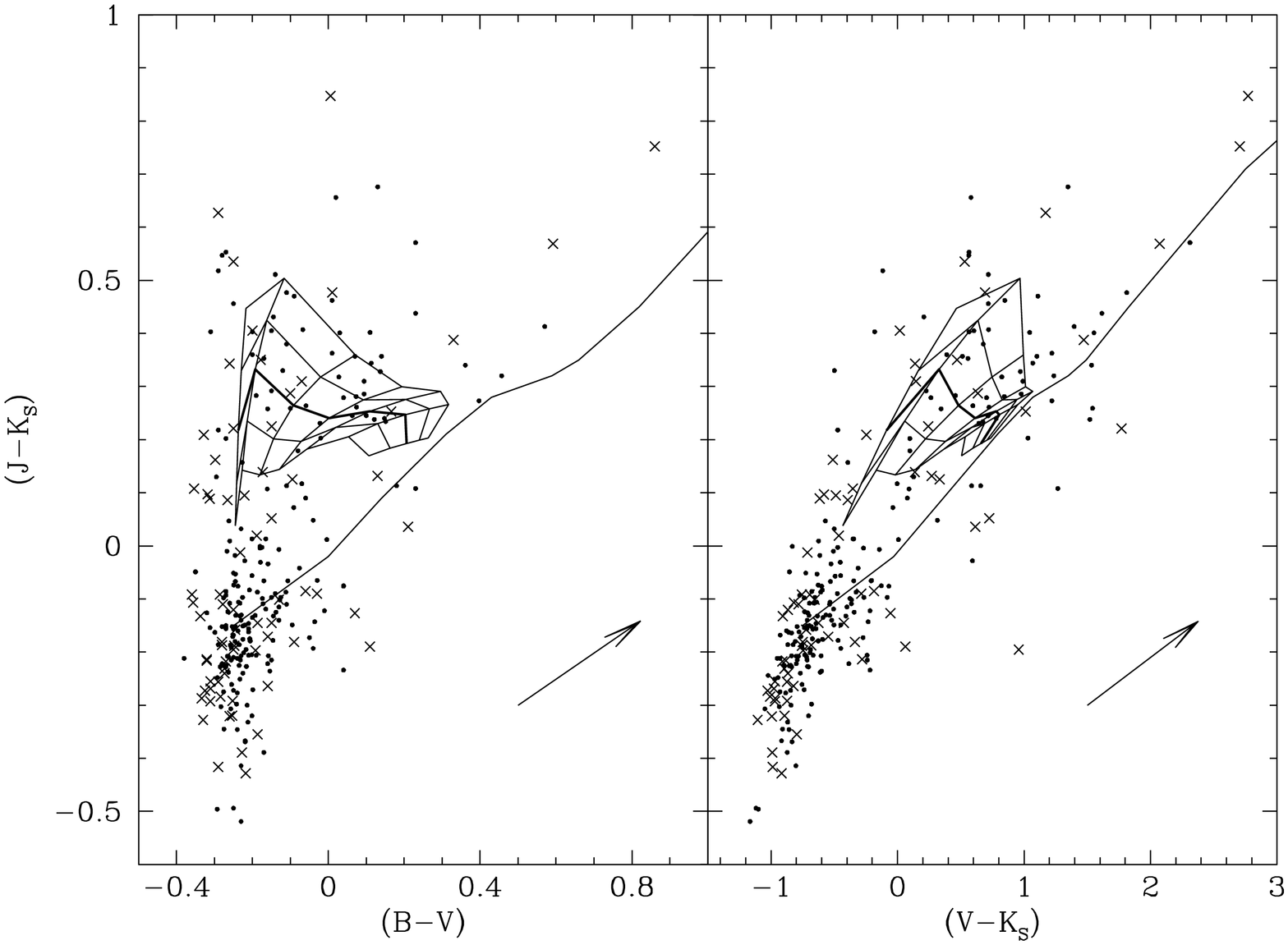}}

  \caption{Same as Figure~\ref{MScolortrack}, except all sdBs
  (circles) and sdOs ($\times$) with 2MASS 2IDR photometry are shown as in
  Figure~\ref{allsdB}. \label{AllMScolortrack} }
\end{figure}

The (main-sequence) loci are plotted with the entire hot
subdwarf sample in Figure~\ref{AllMScolortrack}.

These simple models give results consistent with previous studies
concerning the companion spectral types and their light
contributions. We cannot, however, assume an {\Mv} for the companion
and invert the calculation case-by-case to give reliable individual
values for {\MvsdB}, nor {\it vice versa}, because of the
uncertainties in the measured magnitudes, colors, and reddening.
Additionally there is the expected intrinsic spread in {\MvsdB}
referred to above {\citep[for further discussion see][]{Saffer}}.  For
similar reasons, neither can we infer exact companion spectral types
for individual objects.

Of the 33 currently known composite hot subdwarf systems, essentially
all of the cool companion stars for which a spectral class is known
are of spectral types G or K.  Our models show that composite systems
with G, K, and early M-type companions show the most significant
differences in color from a single sdB (particularly in {\JKs}) and
from main sequence stars.  This could explain why most {\it known}
composite systems have GK-type companions: for companions earlier then
$\sim$G, the combined system looks more like a main sequence star (or
a cool, {\PopII} subdwarf) than a sdB star. Such systems would likely
be excluded from catalogs of hot subdwarfs.  For main-sequence
companions much later than $\sim$M0, the effect on the combined system
light is so small that the companion would not easily be detected in
the visual.  Most of the spectroscopic studies of composite sdBs have
been done in the visual and blue region of the spectrum where very
little color difference between a composite sdB+dM and a single sdB
exists, and there are indeed few identifications of dM companions ---
thus far there are only four known: PG~1241$-084$ {\citep[also
HW~Vir;][]{M+M}}, PG~1336$-018$ {\citep{Kil98}}, HS~0705+6700
{\citep{Drechsel}}, and PG~1017$-086$ {\citep{Maxted02}}.  All of
these are very short period binaries ({2\hour48\minute},
{2\hour25\minute}, {2\hour18\minute}, and {1\hour45\minute}
respectively), and the the nature of the companion was studied in part
by eclipses or reflection effects in the light curve.  PG~1241$-084$
(KHD\nocite{Kilkenny} sequence {\#}605) and PG~1017$-086$ ({\#}407)
are found in the 2MASS 2IDR.  Colors for such close binary stars,
constructed from non-simultaneous measurements of visible and infrared
magnitudes, may be unusual.  The color indices for these two stars are
nevertheless consistent with single hot subdwarfs ({\#}605
PG~1241$-084$: $\JKs=-0.149$, $\bv=-0.332$, $\VKs=-0.110$; {\#}407
PG~1017$-086$: $\JKs=-0.176$, $\bv=-0.225$, $\VKs=-0.417$).  A few
other objects can be noted in the region of color space that, based on
our simple models, would be occupied by sd+dM composites
(Figures~\ref{Allcolortrack} and~\ref{AllMScolortrack}).  Owing to
uncertainties in the measured magnitudes and colors, however, we
cannot clearly distinguish early M companions from late K companions
--- near-IR or IR spectroscopy will be needed to distinguish clearly
between late K and early M-type companions.  Yet it seems clear, based
on our color plots, that there are not great numbers of ``hidden''
early M-type ($\sim$M0) companions to hot subdwarfs.  For main
sequence companions substantially later than $\sim$M0, the {\JKs}
color is rapidly returning to that of a single subdwarf.

Identifying and characterizing further M-type companions to sdB stars
would help settle some of the questions about the nature of
earlier-type (G--K) companions, in particular whether they are
predominately main sequence or subgiant stars.  Late M-type companions
can safely be assumed to be on the main sequence.  If many of these
are found among the sdB stars with ``composite'' colors, the
relatively faint {\MvsdB} inferred from such systems would imply that
the G--K type companions in other composite-color systems are also
close to the main sequence.  If late-M companions are not found in
great numbers, it could be inferred that the G--K companions are
mostly subgiants, since the corresponding bright {\MvsdB} inferred
from this assumption would explain the absence of sd+dM systems
detectable from photometry.  The existence of many subgiant
companions, with their short lifetimes as subgiants, would be
interesting, since the evolutionary timescale of sdB stars is also
short.

\subsection{Hot Subdwarf Distribution in {\JKs} Color} \label{sec:JKs}

\begin{figure}
  \epsscale{1}
  \plotone{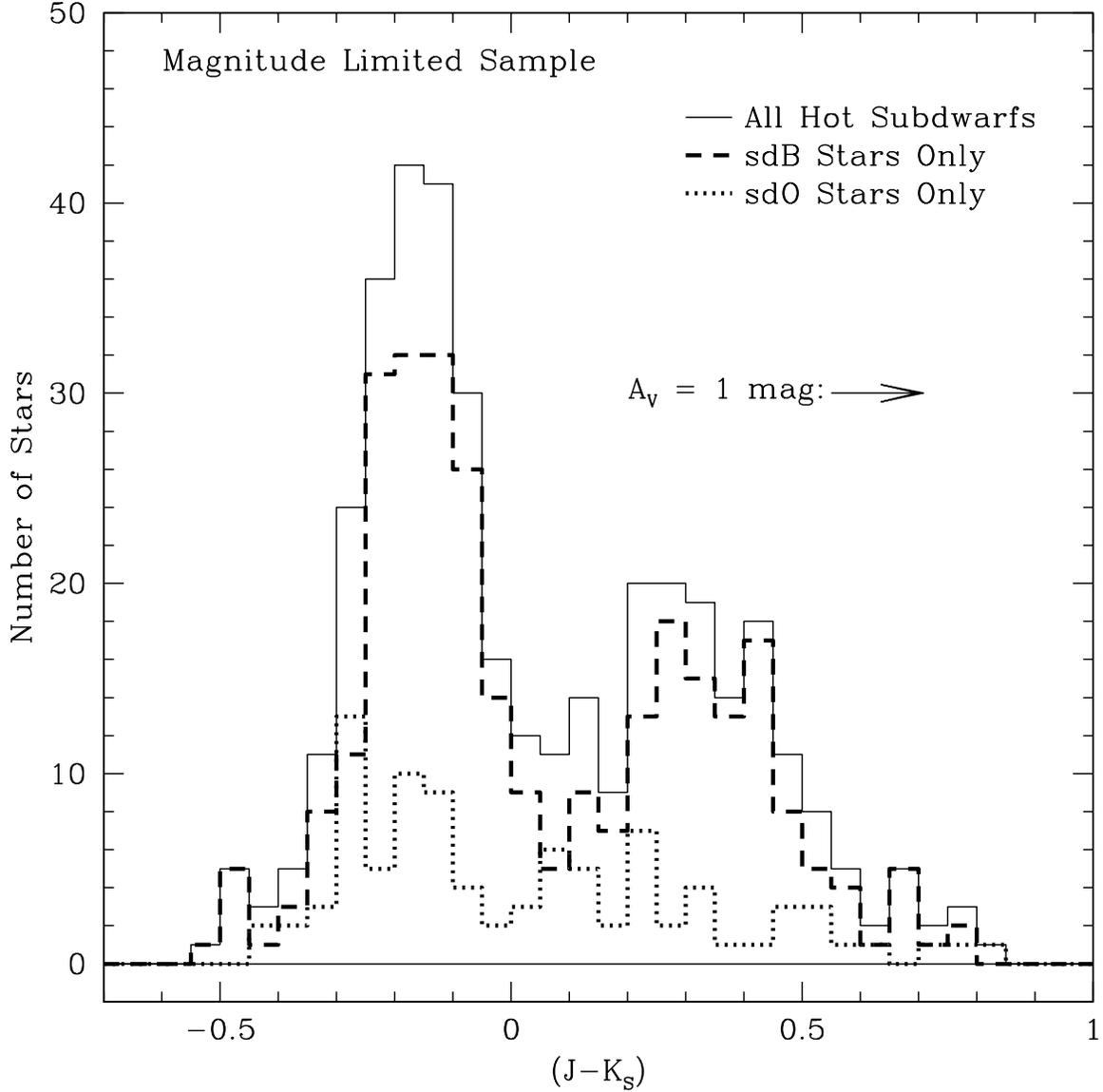}

  \caption{Histogram of {\JKs} for all hot subdwarf stars found in the
  2MASS 2IDR (bin size 0.05~mag).  The solid line indicates the full sample
  of hot subdwarfs, the dashed line indicates the sdB group, and the
  dotted line indicates the sdO group (these groups were defined in
  {\S\ref{sec:Sample}}).  Single subdwarfs are believed to reside in
  the left peak of the distribution centered around $\JKs \approx
  -0.15$, while composite systems (sd+later type) reside in the right
  peak centered around $\JKs \approx +0.3$, with the sdOs falling to
  slightly bluer colors than the sdBs in general.  
  A tabulation of the numbers and fractions of single and composite
  hot subdwarfs can be found in Table~\ref{Fractions}.
  The arrow indicates the shift in {\JKs} color due to one magnitude
  of extinction at $V$.  \label{JKhist} }
\end{figure}

Since visual photometry is not available for all of the hot subdwarfs
contained in the 2IDR, and since there is a more distinct separation
between composites and single subdwarfs stars in {\JKs} than in the
other colors, we examined the distribution of subdwarfs based on their
{\JKs} colors alone. A histogram of {\JKs} for all subdwarfs with $J$
and {\Ks} measurements (385 subdwarfs: 294 sdB, 91 sdO), reveals a
bimodal distribution (Figure~\ref{JKhist}). One peak is centered near
{$\JKs \approx -0.15$} ($\sim$B2--B3 spectral type), and the other is
centered near {$\JKs \approx +0.30$} ($\sim$G5 spectral type).  There
are also wings stretching to both redder and bluer colors.  This
distribution cannot be explained by reddening, because no significant
trends were observed with galactic latitude.  We would also expect
reddening to produce a tail to the red from a single peak, but not a
pronounced second peak.  Finally, the amount of extinction that would
be required is also excessive.

We sorted the sdB and sdO samples into three groups by $J$
magnitude. The bimodal peaks are very pronounced in the brightest
third of each sample, less distinct in the middle third, and
practically non-discernible in the faintest third.  This pattern can
be fully explained by the increase in photometric uncertainty at
fainter magnitudes.

In keeping with the results found from our modeling of color--color
distributions, we postulate that the stars within the blue peak of the
{\JKs} distribution ($\JKs < +0.05$) are single stars (or sd+WD pairs,
with colors indistinguishable from single hot subdwarfs), and that
stars in the red peak ($\JKs > +0.05$) are composite (sd+late type)
systems.  The color index separating these two groups of stars was
adopted from the box defining the region of single subdwarfs in
Figure~\ref{Known}.  With this definition, we find that 58$\pm$5{\%}
of the subdwarf stars in our 2MASS 2IDR sample are apparently single,
and 42$\pm$5{\%} are composite.  The quoted 95{\%} confidence errors
are sampling errors computed from the binomial distribution, taking
the size of the sample into account.  If the color index of the
boundary separating single and composites is increased (decreased) by
0.05 mag, the fraction of subdwarfs that are ``blue'' (i.e., single)
is increased (decreased) by 3{\%} (3\%).

While the 2IDR sample is representative of the KHD\nocite{Kilkenny}
catalog as a whole, it is likely that neither of these is
representative of the true hot subdwarf population in the solar
neighborhood.  There are the usual completeness issues, including
heterogeneous magnitude limits and different classification criteria
used by the sources from which KHD\nocite{Kilkenny} was compiled
{\citep[although we note that about 2/3 of the cataloged subdwarfs are
drawn from the PG survey,][which has rather well-defined
properties]{PG}}.  We draw renewed attention, however, to a kind of
censorship bias that is present in the PG survey. An unknown number of
composite subdwarfs may have been excluded ``by design'' (i.e., not
included in the PG catalog and never identified as hot subdwarfs),
since according to {\citet{PG}} UV excess stars were omitted from
their catalog if their spectra showed a pronounced {\CaII}~K
line. (The supposition was that many of these would be metal-weak F or
G subdwarfs, which met the purely photometric criterion for inclusion
as UV excess stars owing to a combination of reduced line blanketing
and larger than average photometric errors.)  Thus composite hot
subdwarf systems with F or G companions might be {\it
underrepresented} in KHD\nocite{Kilkenny} and the present sub-sample
drawn from it.  (On the other hand, since some PG subdwarfs are found
in other independent surveys, some of the composite systems
``missing'' from PG may nevertheless have been included in the
KHD\nocite{Kilkenny} compilation, from other sources.)  The number of
``excluded'' objects is roughly 1000--1200 (R.~F.~Green 2003, private
communication), but the fraction of these that may be composite hot
subdwarfs rather than G subdwarfs is not known, so these selection
effects are difficult to quantify.  In time, objective prism-based
surveys such as the one being carried out with the Hamburg Schmidt
telescope may provide a catalog of hot subdwarfs that is more clearly
representative of the Galactic population.  Meanwhile, we proceed to
analyze the 2MASS 2IDR sample from the KHD\nocite{Kilkenny} catalog.
It must be realized, however, that since the 2MASS 2IDR represents a
magnitude limited survey, the composite systems (being more luminous
than single subdwarfs) will be {\it overrepresented} in this sample,
other things being equal (see {\S\ref{sec:Vol}} for further discussion
of this point).

The same bimodal distribution in {\JKs} exists in the sub-sample of
sdB/sd stars that is found in the full sample (58$\pm$6{\%} have $\JKs
< +0.05$, and 42$\pm$6{\%} have $\JKs > +0.05$).  The sub-sample of
sdO/sdOB stars shows a less clear pattern. There is a hint of three
groups: one near $\JKs \sim -0.2$, another near $\JKs \sim +0.2$, and
the third near $\JKs \sim +0.5$, but the small number of stars in each
group makes this finding quite tentative.  The two larger groups of
sdO stars ($\JKs \sim -0.2$ and +0.2), if real, can be considered to
fall to the ``blue'' side (lower {\JKs}) of the the two peaks in the
sdB distribution, consistent with the higher temperatures of the sdO
with respect to the sdB stars.  Overall, the observed bimodal
distribution in {\JKs} of all hot subdwarfs is defined by the majority
sdB stars.  The fraction in each peak of the {\JKs} distribution does
not significantly change if the sdO stars are removed.  A breakdown of
the fractions of subdwarfs by {\JKs} color is shown in
Table~\ref{Fractions}.

\subsection{Fitting the sdB {\JKs} Distribution} \label{sec:Gauss}

\begin{figure}
  \epsscale{1}
  \plotone{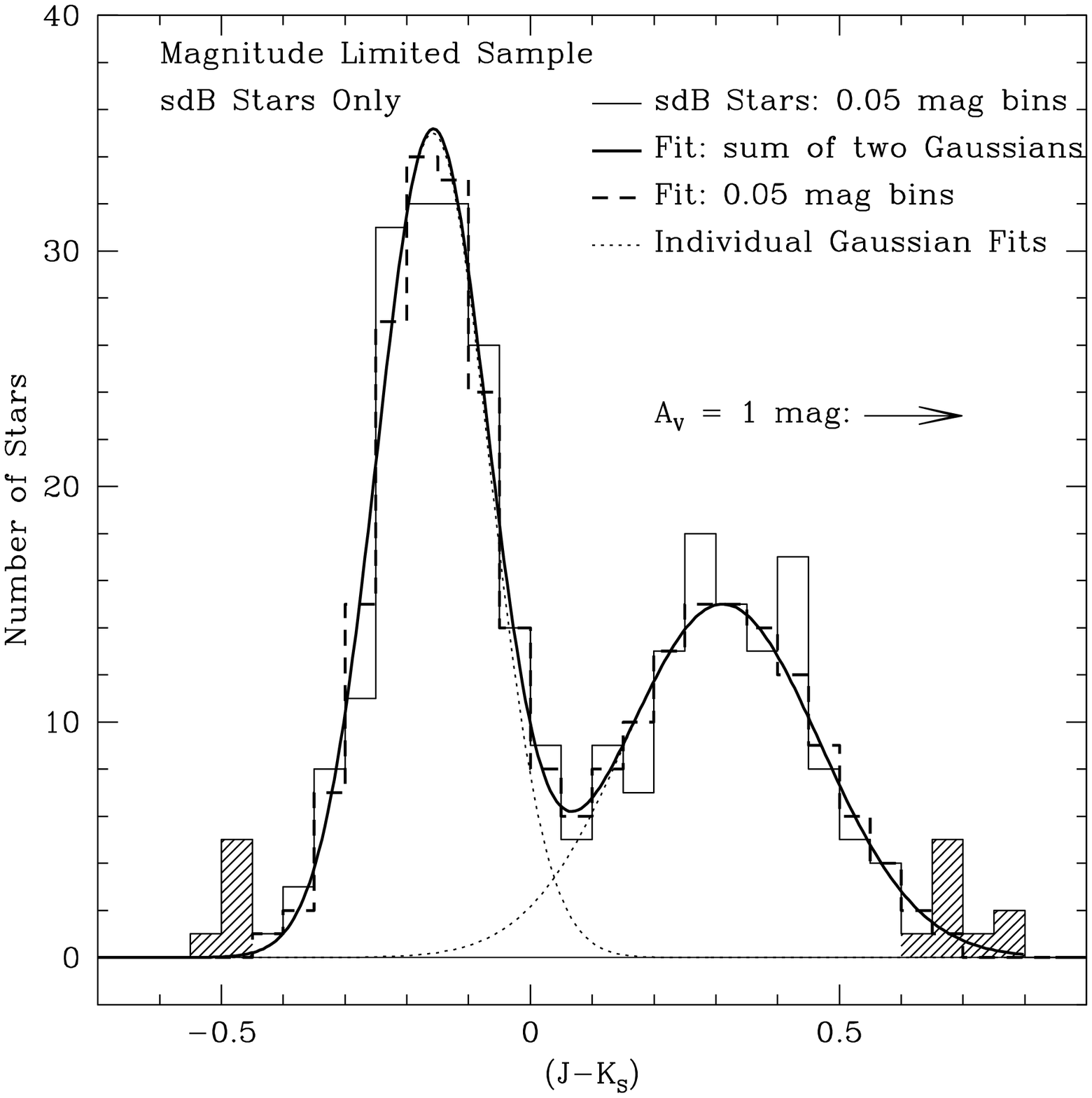}

  \caption{The sdB only distribution in {\JKs} (from
  Figure~\ref{JKhist}) fitted by the sum of two Gaussians.  Shaded
  bins were excluded when calculating {\chisq} and {\chisqr}.  See
  {\S\ref{sec:Gauss}} for further discussion, and
  Table~\ref{GaussParameters} for a tabulation of the fit parameters.
  \label{GaussAll} }
\end{figure}

We have fitted the bimodal {\JKs} distribution of the sdB stars
(alone) to the sum of two Gaussians, centered on the blue and red
peaks.  An initial ``fit-by-eye'' to the binned data was improved by
minimizing {\chisq} on a range that was trimmed ($-0.45 < \JKs <
+0.6$) to remove the effects of the reddest and bluest outliers.  The
{\chisq} minimization was done using a systematic gridded search in
the neighborhood of the initial fit.  There are six fitting
parameters: two mean values (or centers), two dispersions (or widths),
and two normalizations (or amplitudes).  The final fit has $\chisq =
6.8$ (reduced $\chisqr = 0.45$, with 15 degrees of freedom).  The
parameters for the fit are listed in Table~\ref{GaussParameters} and
the fit is shown in Figure~\ref{GaussAll}.

While the {\chisq} statistic provides a useful criterion for improving
the fit, its final numerical value should be used with caution, since
the range of the fit was trimmed, and since some of the data bins
contain only a few members.  In particular, the 95{\%} confidence
intervals shown in Table~\ref{GaussParameters} are indicative rather
than definitive.  The small value of {\chisqr} nevertheless
demonstrates the plausibility of a two-Gaussian fit, which is
confirmed by visual inspection.  The ``blue'' Gaussian has a somewhat
smaller Gaussian dispersion than the red one, but both dispersions
($\sigma \approx 0.09$, 0.16) are close to the average photometric
error for this sample from 2MASS, $\sigma(\JKs) \approx 0.12$.  This
makes it difficult to state whether there is any intrinsic spread in
the {\JKs} colors of the single or composite sdBs.

To the extent that the separation of single and composite sdB stars in
the sample is accurately represented by two Gaussians, it is seen that
there is very little overlap in {\JKs} color index between the two
groups of stars.  The areas of the two Gaussians are in the proportion
57:43, consistent with the binomial result found using a sharp
demarcation at $\JKs = 0.05$.

\section{A ``Volume Limited'' Hot Subdwarf Sample} \label{sec:Vol}

\subsection{Defining the Sample} \label{sec:DefineVol}

There is a strong advantage in working with the {\JKs} color index to
discuss the fraction of composite-color binaries among hot subdwarfs,
as in {\S\ref{sec:JKs}} and {\ref{sec:Gauss}} above, since the data
are available for a larger sample than would be the case for optical
or mixed optical-infrared colors.  Also, the $J$ and {\Ks} from 2MASS
are homogeneous in nature.  However, a strong selection bias in these
data exists, because 2MASS itself is an approximately {\it magnitude}
limited survey (albeit with different magnitude limits in each of
three passbands).  The entire sample of hot subdwarfs we have been
discussing (common to both KHD\nocite{Kilkenny} and 2MASS 2IDR) is
thus a {\it magnitude} limited sample in some sense.  The {\it volume}
sampled for single subdwarfs is therefore smaller than that sampled
for composites, since the presence of the late type companion
increases the distance at which a composite system remains brighter
than the limiting 2MASS magnitude.

To further explore this selection bias, we constructed ($J$, {\JKs})
and ({\Ks}, {\JKs}) color-magnitude diagrams for the full 2MASS 2IDR
sample (Figure~\ref{VolCut}).  Several interesting patterns are noted.
First, the two populations (single and composite) can be seen, and
both are observed to cover the entire magnitude range (i.e.,
composites and/or singles are not ``clumped'' at a specific apparent
magnitude).  Second, a majority of the objects with extreme {\JKs}
colors (either blue or red) are found at the faintest magnitudes, an
effect largely explained by photometric errors.  Third, it can be seen
that the sample is limited by detections in the {\Ks} band. There is a
clear cut-off at $\Ks \approx 15.75$, while in $J$ the cut-off
magnitude varies by color (redder objects that are detected in {\it
both} $J$ and {\Ks} run to a fainter $J$ magnitude than the bluer
objects).

The interpretation of this third feature is that a blue (single)
object detected near $J = 15.5$ or fainter is missing at {\Ks} and
thus is not part of the plotted sample; a red (composite) object at
the same $J$ will also have a {\Ks} detection and thus will appear in
Figure~\ref{VolCut}.  (This effect is distinct from the fact that
composites appear brighter at $J$ due to the combined light from two
stars.)

Recognizing this selection bias, we can attempt to define an
approximately {\it volume} limited sample of hot subdwarfs.  (To our
knowledge, this is the first time that this bias has been discussed in
the context of hot subdwarf composites.  {\citet{TMSH}} discussed a
``statistically complete'' sample containing 11 hot subdwarfs, however
they did not address the composite vs.\ single bias.)  To define our
volume limited sample, we try to remove those subdwarfs that we
observe to be composite, but which would {\it not} have been detected
by 2MASS had they been single.  With only a single color index to work
with, and with uncertainties in both the photometry and the relative
luminosities of the hot subdwarfs and their companions, we cannot rely
on such a technique to say whether any given individual system lies
within the defined volume.  The sample so constructed can be used in a
statistical sense, however, to reflect the general nature of the true
distribution, and at least to indicate the size of the selection
effect.

\begin{figure}
  \epsscale{0.77}
  \rotatebox{-90}{\plotone{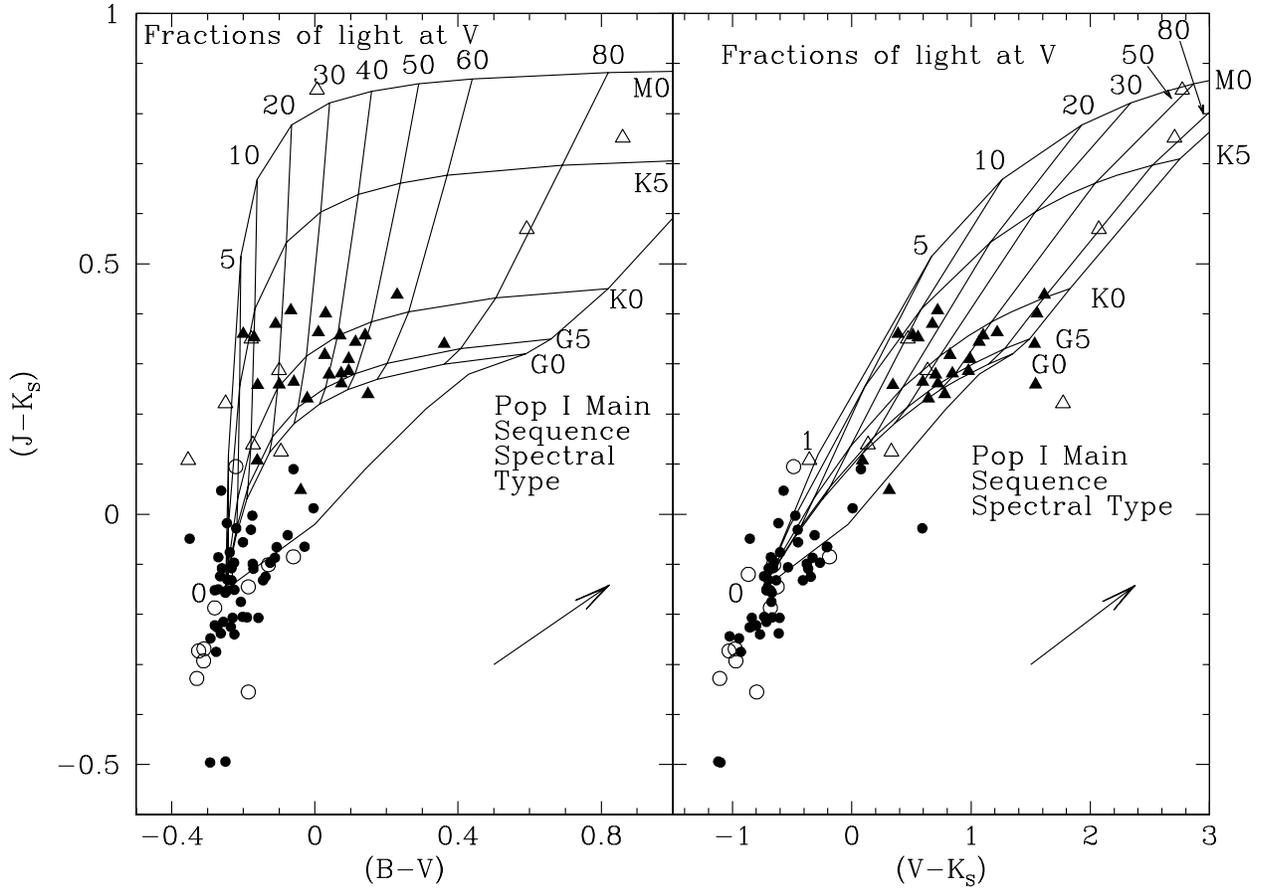}}

  \caption{Grid of composite colors computed by combining the light
  from a hot subdwarf ($\bv \approx -0.25$, $\VKs \approx -0.75$,
  $\JKs \approx -0.15$) with that of ``typical'' {\PopI} main
  sequence stars {\citep[G0, G5, K0, K5, and M0;][]{Johnson}}
  assuming varying fractional contributions to the total light in the
  {\it V} band by the late type star (fractions contributed by the
  companion: 0, 1, 5, 10, 20, 30, 40, 50, 60, 80, and 100{\%}).
  Circles are the reported single subdwarfs (solid: sdB, open: sdO)
  and triangles are the reported composite subdwarfs (solid: sdB,
  open: sdO) from Figure~\ref{Known}.  The large arrows indicate the
  color change due to one magnitude of extinction at
  $V$. \label{colortrack} }
\end{figure}

\begin{figure}
  \epsscale{0.77}
  \rotatebox{-90}{\plotone{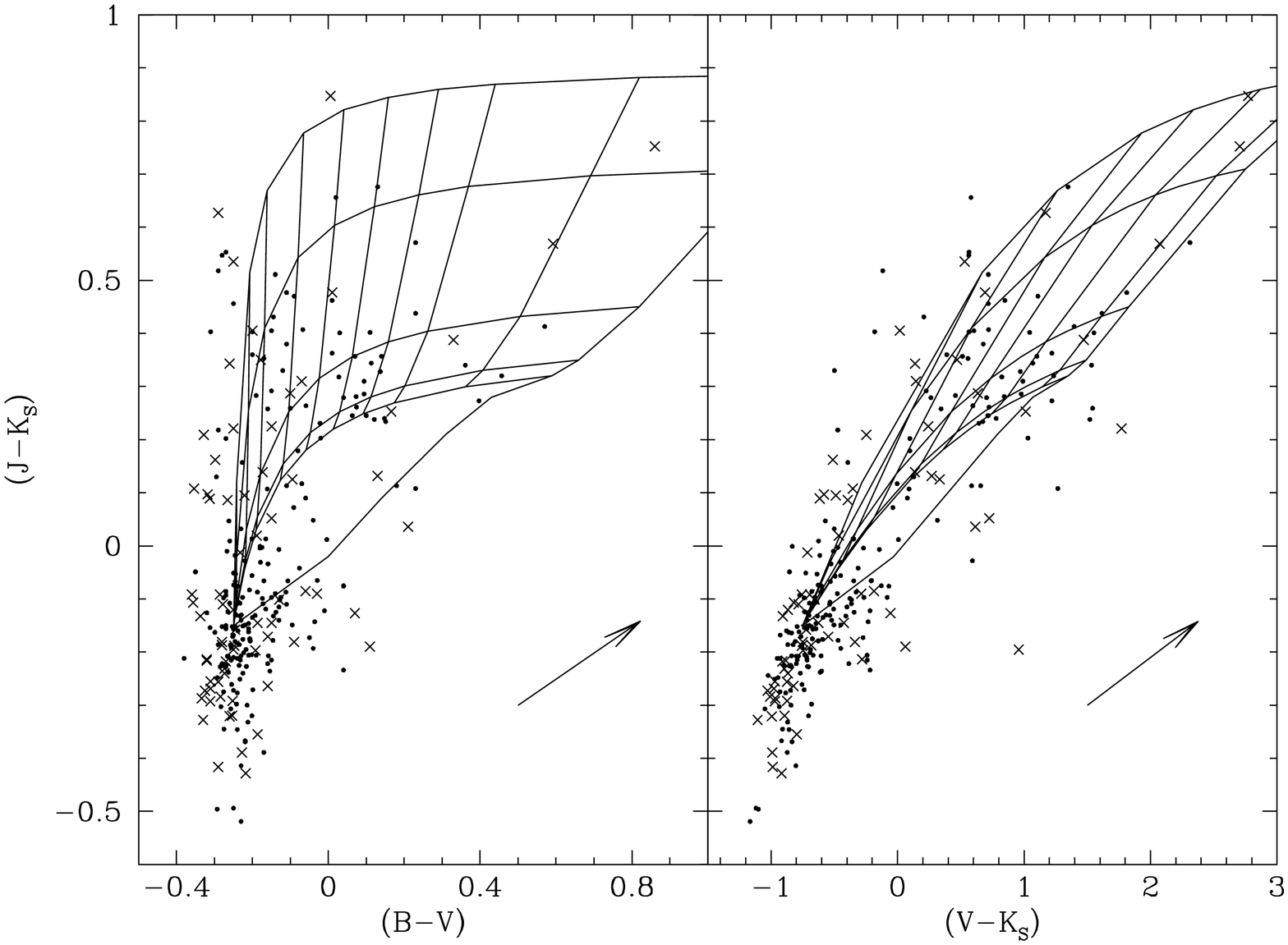}}

  \caption{Same as Figure~\ref{colortrack}, except all sdBs (circles)
  and sdOs ($\times$) with 2MASS 2IDR photometry are shown, as in
  Figure~\ref{allsdB}. \label{Allcolortrack} }
\end{figure}

\begin{figure}
  \epsscale{1}
  \plotone{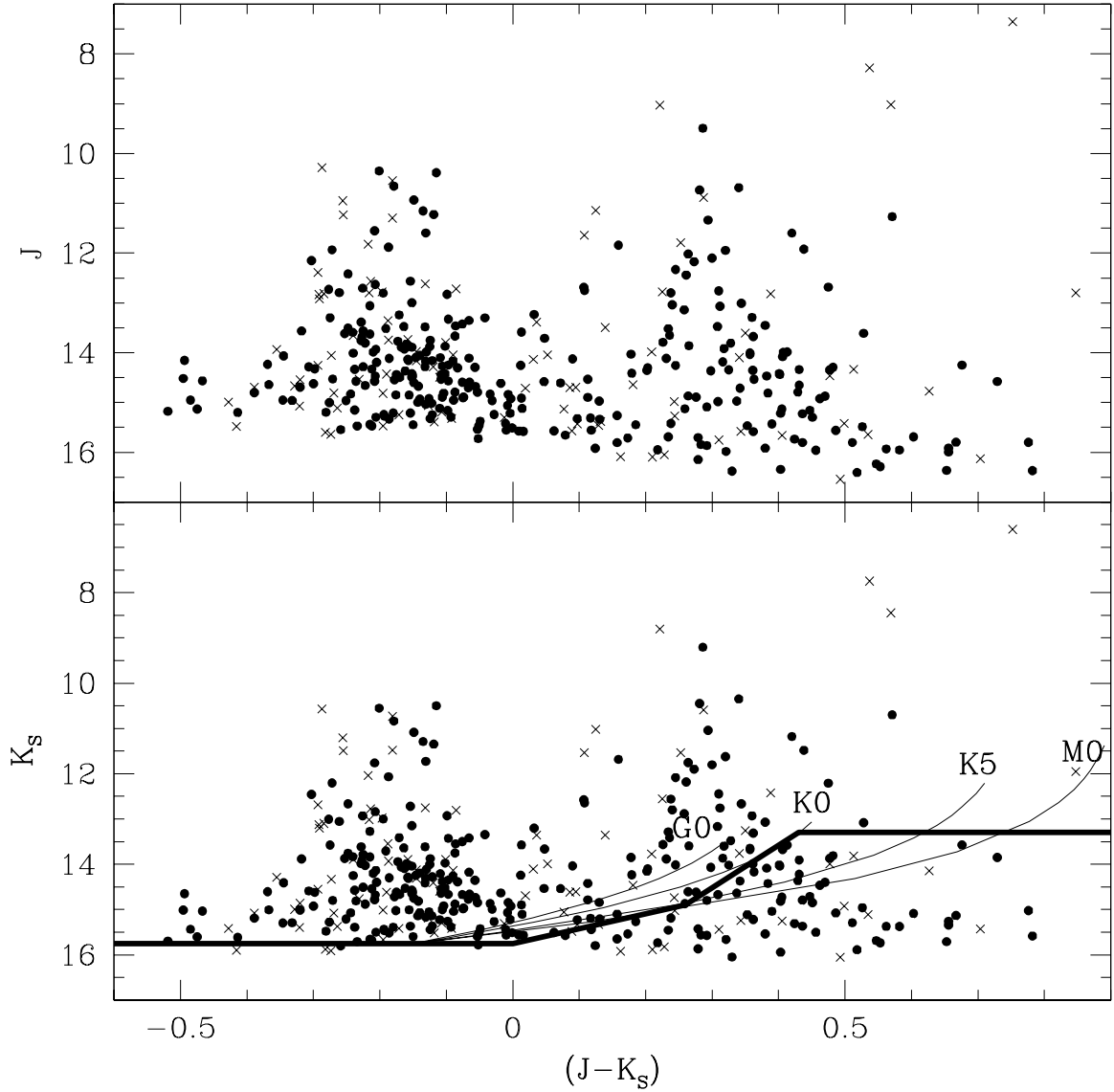}

  \caption{Color-magnitude diagrams of $J$ (top) and {\Ks} (bottom)
  vs.~({\JKs}). The sdBs are plotted as circles and the sdOs as
  crosses (``$\times$'').  The {\Ks} vs.~({\JKs}) diagram shows the
  cut made to get an approximately volume limited sample (thick line),
  as well as the limits (thin lines) for companions with colors
  corresponding to the {\PopI} main sequence spectral types: G0, K0,
  K5, and M0 (these limits are the lines from Figure~\ref{colortrack}
  transformed into color-magnitude space by assuming an apparent magnitude
  for the subdwarf of $\Ks = 15.75$).  \label{VolCut} }
\end{figure}

The observed cut-off magnitude from Figure~\ref{VolCut} for a single
hot subdwarf is $\Ks \approx 15.75$.  Taking this limit as the
apparent magnitude of the hot subdwarf component in a composite
system, we can compute the corresponding apparent magnitude for the
total light from the system, using the models discussed in
{\S\ref{sec:Model}} and shown in Figures~\ref{colortrack}
and~\ref{Allcolortrack}.  This total magnitude will vary as the
spectral type and/or relative luminosity of the late type companion is
varied.  The locus of thus-computed magnitudes is shown in the lower
panel of Figure~\ref{VolCut} for a hot subdwarf with the ``typical''
color $\JKs = -0.15$ adopted in {\S\ref{sec:Model}}, for companions
with colors of the {\PopI} main sequence spectral types G0, K0, K5,
and M0.  Along each of these lines, the color and total magnitude
varies as the relative brightnesses of the two component stars are
varied.  Note that {\it no assumption has been made about the absolute
magnitude of either the hot subdwarf or the cool companion.}

We now replace the family of lines (corresponding to different
spectral types of the cool star) with a single representative line.
(This is because based on {\JKs} alone we cannot know the true
spectral types of the late type companions without making assumptions
about the relative luminosities of the component stars.)  We use this
line to approximate the appropriate magnitude limit for a volume
limited sample, again in a {\it statistical} sense.  The line is
chosen to pass though the middle of the known composite sdBs shown in
Figure~\ref{colortrack}; it thus represents a $\sim5${\%}~contribution
(at $V$) by a K5 star, or a contribution of 40--80{\%} at K0 in
Figure~\ref{VolCut}.  The change in the volume-limiting {\Ks}
magnitude is thus most rapid near $\JKs \approx +0.3$ (see
Figure~\ref{colortrack}).  All hot subdwarf stars in the ({\Ks},
{\JKs}) color-magnitude diagram that fall above this line are included
in our statistical volume limited sample.  The resulting sample size
is 300 hot subdwarfs (228 sdB and 72 sdO).  The average error in
{\JKs} for this new sample is 0.10 mag (slightly reduced from 0.12
mag, which applies to the full {\JKs} sample).

\subsection{Color-Color and {\JKs} Distribution of the Volume
 Limited Sample} \label{sec:VolColor}

\begin{figure}
  \epsscale{0.77}
  \rotatebox{-90}{\plotone{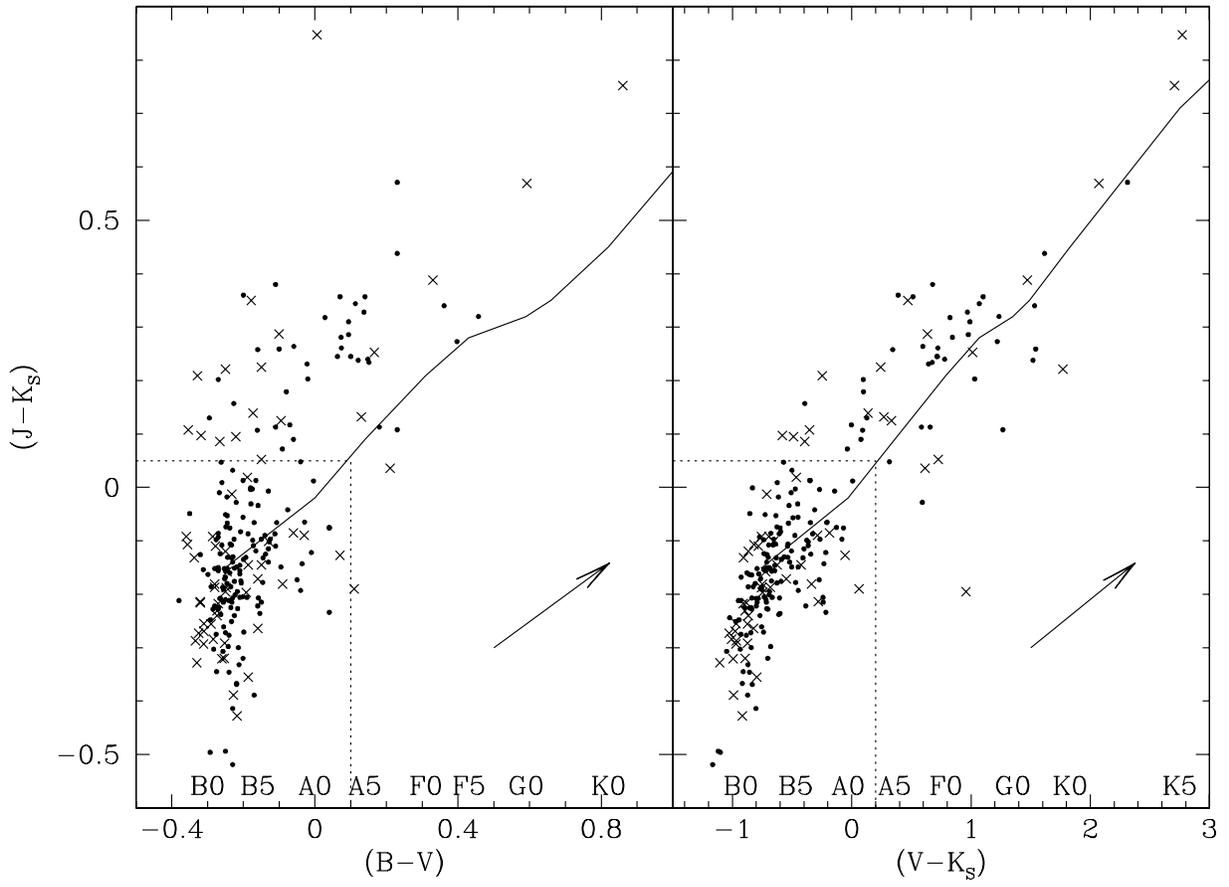}}

  \caption{Same as Figure~\ref{allsdB}, except displaying only hot
  subdwarfs in the ``volume limited'' sample; sdBs (176 in the
  left panel, 180 in the right) are denoted with circles and sdOs
  (62 in both panels) with crosses (``$\times$'').
  \label{trimmedcolor} }
\end{figure}

For the volume limited sample, visual magnitudes ({\BV} or the {\uvby}
equivalent) are available for 176 sdB stars (4 additional with $V$
only) and 61 sdO stars (1 additional with $V$ only).  The volume
limited sample occupies the same regions of ({\JKs}, {\bv}) and
({\JKs}, {\VKs}) color spaces (Figure~\ref{trimmedcolor}) that the
magnitude limited sample does (compare Figures~\ref{allsdB}
and~\ref{trimmedcolor}), but many of the more extreme red objects (in
{\JKs}) have been removed.

\begin{figure}
  \epsscale{1}
  \plotone{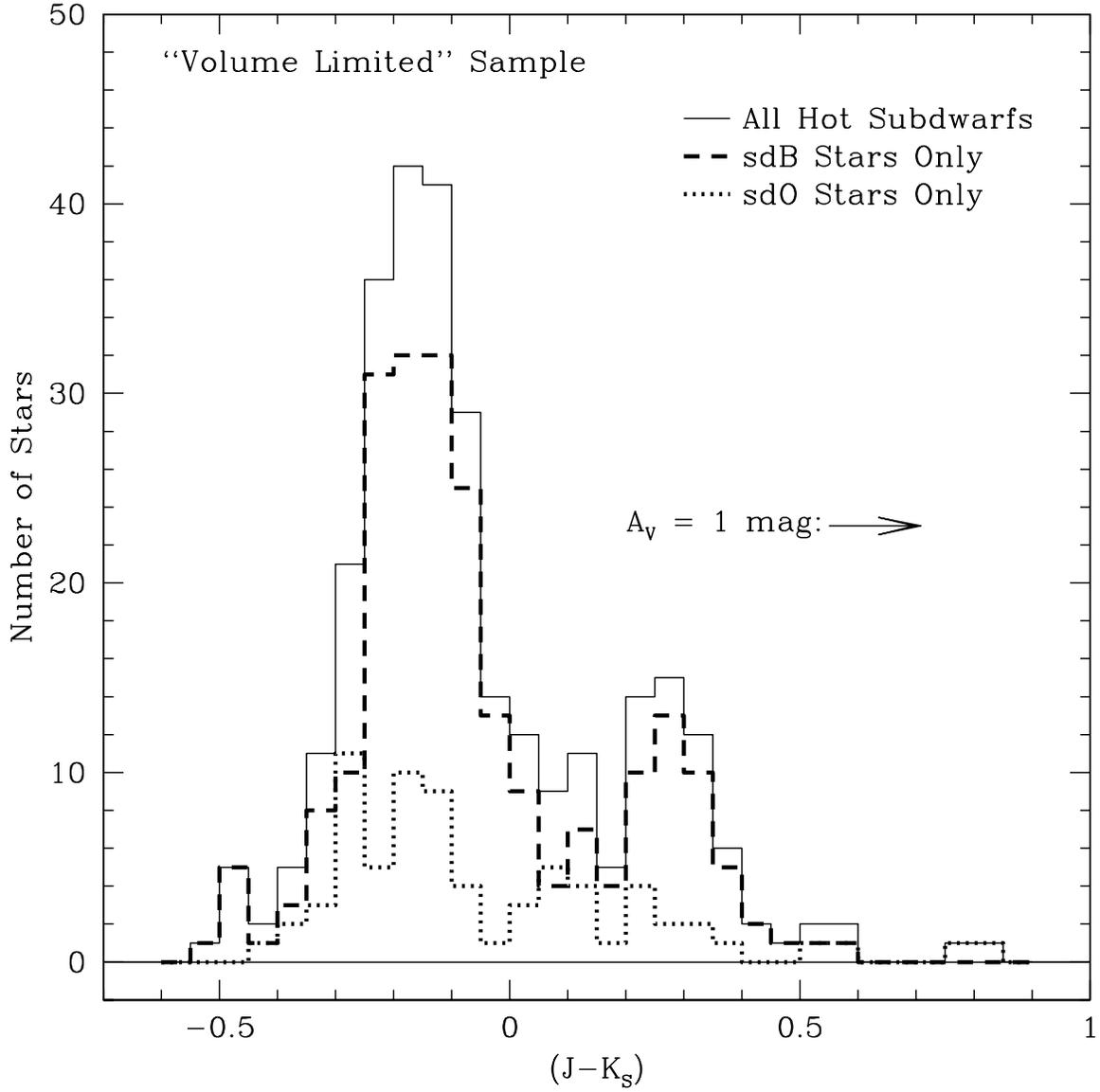}

  \caption{Same as Figure~\ref{JKhist}, except showing only hot
  subdwarfs in the ``volume limited'' sample.
  A tabulation of the numbers and fractions of single and composite
  hot subdwarfs can be found in Table~\ref{VolFractions}.
  The arrow indicates the shift in {\JKs} color due to
  one magnitude of extinction at $V$. \label{VolHist} }
\end{figure}

The {\JKs} histogram made using only the volume limited sample
(Figure~\ref{VolHist}) can be compared with the corresponding diagram
for the entire magnitude limited sample (Figure~\ref{JKhist}).  The
distribution is still bimodal, with one peak centered near $\JKs
\approx -0.15$ as before ($\sim$B2--B3 spectral type), and the other
near $\JKs \approx +0.30$ ($\sim$G5 spectral type).  The sdB stars
still dominate the total hot subdwarf distribution, and the sdO stars
show a stronger hint of bimodality (again with both peaks shifted
slightly blueward of the sdB peaks).  The blue wing to the
distribution remains, while the red wing has been greatly reduced.
Most important, the fractions of single (``blue'') and composite
(``red'') systems have been altered (Table~\ref{VolFractions}).
Whereas the magnitude limited sample showed a 58:42 split between
single and composite sdB stars (Table~\ref{Fractions}), the volume
limited sample shows a 75:25 split in favor of single sdB stars.
Likewise, the split for the total hot subdwarf sample is 73:27 in the
volume limited sample (as opposed to 58:42 in the magnitude limited
sample).

\begin{figure}
  \epsscale{1}
  \plotone{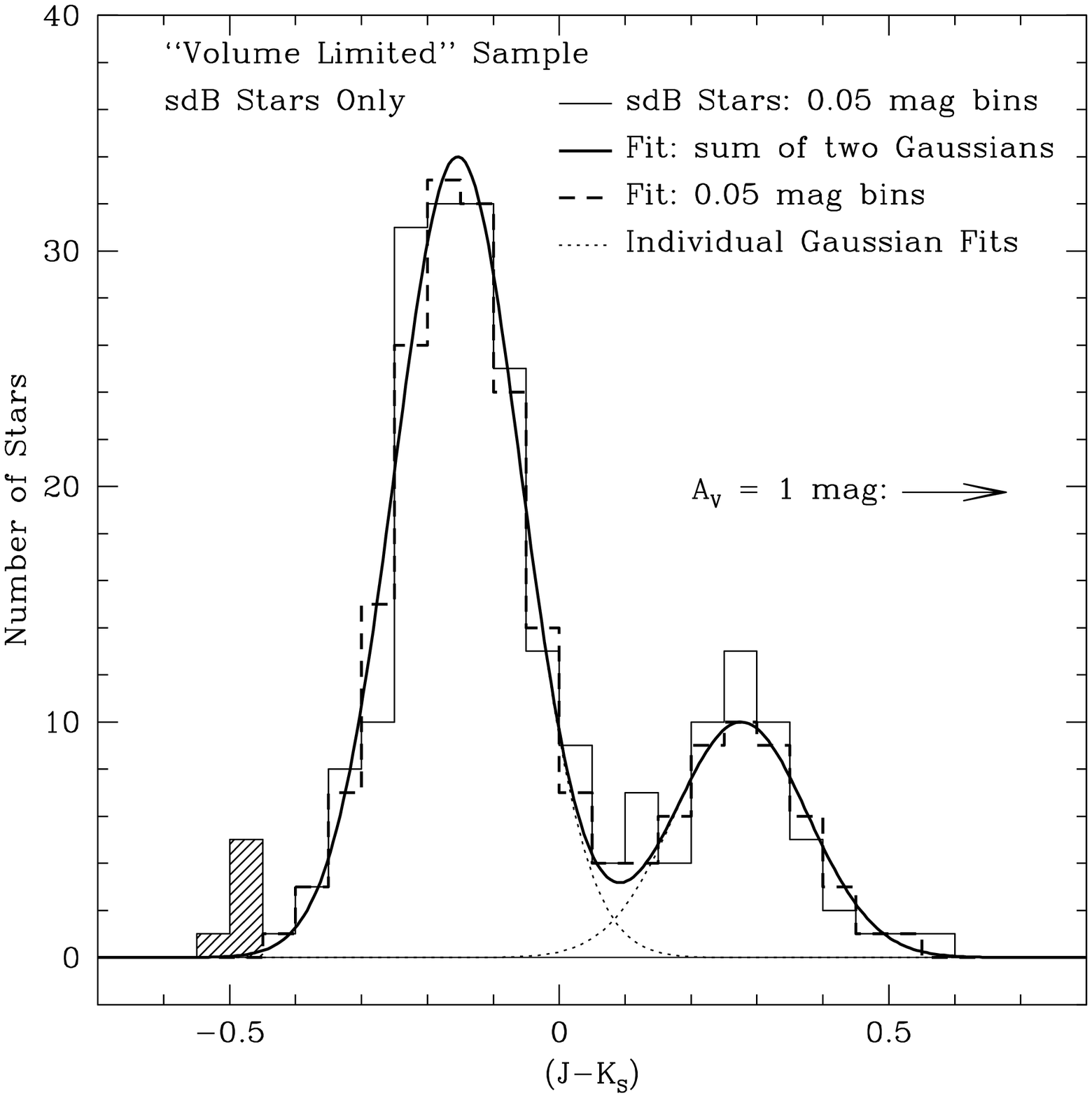}

  \caption{The ``volume limited'' sdB-only distribution in {\JKs}
  (from Figure~\ref{VolHist}), fitted by the sum of two Gaussians.
  Shaded bins were excluded when calculating {\chisq} and {\chisqr}.
  See {\S\ref{sec:VolColor}} for further discussion, and
  Table~\ref{VolGaussParameters} for a tabulation of the fit parameters.
  \label{Gauss} }
\end{figure}

We can again describe the sdB {\JKs} distribution for the volume
limited sample as the sum of two Gaussians.  The new $\chisq = 8.9$
(reduced $\chisqr = 0.59$, with 15 degrees of freedom).  The
parameters for the fit are given in Table~\ref{VolGaussParameters},
and the fit is shown in Figure~\ref{Gauss}.

The parameters of the blue Gaussian fits for the magnitude limited and
volume limited samples are essentially the same.  This follows because
the number of stars that fall in the blue peak was not substantially
altered in creating the volume limited sample.  The width of the blue
Gaussian is consistent with the average photometric error from 2MASS
for this sample, $\sigma(\JKs) \approx 0.10$, implying that there is
little intrinsic spread in the {\JKs} color index of single sdB stars.

The red Gaussians for the magnitude limited and volume limited samples
have different mean values (centers), with the volume limited sample
shifted blueward relative to the magnitude limited sample
[$\Delta(\JKs) \approx -0.04$]. The normalizations (amplitudes) are
different, owing to to the decrease in the number of red stars in the
volume limited sample.  The change in center is likely due to the
removal of the reddest and also the faintest (and thus most uncertain)
measurements from the red peak.  The dispersion of the red Gaussian
that represents the volume limited sample, $\sigma(\JKs) \approx0.10$,
is reduced from the magnitude limited case, and is now consistent with
the 2MASS photometric error for this new sample ($\approx0.10$ mag);
this implies that there is little variation in the {\JKs} color of
the remaining composite systems.

\section{Summary} \label{sec:Summary}

Using updated and improved coordinates for hot subdwarfs cataloged by
KHD\nocite{Kilkenny}, we have retrieved 2MASS 2IDR photometry for 571
hot subdwarfs.  This is the largest sample of hot subdwarfs, and in
particular composite systems, ever examined.  We divided the sample
into two sub-samples based on their classifications in
KHD\nocite{Kilkenny}: sdB/sd (``sdB'' group), sdO/sdOB
(``sdO'' group).  We examined these two sub-samples both independently
and combined as a whole.  We examined the distribution of hot
subdwarfs in optical-IR color-color plots, by combining 2MASS with
visual photometry ({\BV} or {\uvby}) from KHD\nocite{Kilkenny} (or a
few selected other sources).  The most revealing combinations of color
indices we found were: ({\JKs}, {\bv}) and ({\JKs}, {\VKs}).  We also
examined the distribution of hot subdwarfs in {\JKs} only and found a
bimodally distributed population.  Finally, we attempted to define an
approximately volume limited sample of hot subdwarfs for statistical
purposes.  This is the first time (to our knowledge), that the
selection bias favoring composite hot subdwarfs in magnitude limited
surveys has been discussed, and an approximately volume limited
statistical sample defined.

In keeping with the exploratory nature of this work and the large
sample size, we have favored a statistical approach, in contrast to
the attempt to find a best characterization of individual binaries.
(The relatively large mean errors of the 2MASS photometry also figured
in the decision to use this approach.)  In the same spirit, we have
contented ourselves with ``naive'' modeling of the colors, e.g.,
adopting single values for the colors and magnitudes of the subdwarfs
rather than a distribution of these quantities --- our goal in the
modeling is to support the overall division of the sample of stars
into single and composite groups, by demonstrating the plausibility of
such modeling.

We found that the locations of hot subdwarfs in visual-IR color-color
diagrams can be modeled (very simply and without assuming luminosities
for either the hot subdwarf or the companion) as either a single
B-type star or as a combination of two stars (spectral type B
representing the hot subdwarf, and spectral type G--K for a companion)
with contributions from the late type companion of $\sim10-60${\%} at
$V$.  To a first approximation sdO stars can be described in the same
manner as sdB stars.  If main sequence companions are adopted, then
the sdB stars are best described with $\MvsdB \approx 4.5-5.0$~mag,
and they typically have G--K type companions.  However, the observed
distribution can be reproduced equally well by assuming subgiant
companions with more luminous sdB stars ($\MvsdB
\approx 2.5-3.0$~mag) --- our current data cannot distinguish
between these two possibilities.  In a histogram of the IR color index
{\JKs}, the two peaks of the bimodal distribution can be understood as
single stars (blue peak at $\JKs = -0.158$) and composite systems
(red peak at $\JKs = +0.311$).  This bimodal distribution is also
present in the approximately volume limited sample, again with the two
peaks at $\JKs = -0.154$ and +0.276.  Making a cut at $\JKs =
+0.05$, we found the ratio between single and composite sdB stars is
58:42 (favoring singles) in the magnitude limited sample, while in the
volume limited sample it is 75:25 (favoring singles).  Note that these
ratios {\it do not include any extrapolation to companions that do not
affect the photometry} (i.e., WD or late M~dwarfs), since such
extrapolation depends on the assumptions made about the absolute
magnitude of the subdwarfs or their companions.  The {\JKs}
distribution for both the magnitude and volume samples can be
described as the sum of two Gaussians with very little overlap and
with fractions of single vs.\ composite in agreement with a sharp cut
at $\JKs = +0.05$.  An interpretation of these ratios is difficult due
to biases in the initial subdwarf catalog.

Comparison to other surveys of composite systems is difficult due to
the unknown selection effects involved.  Other surveys have generally
concluded that $\gtrsim50${\%} of hot subdwarfs have late type
companions {\citep[e.g.,][]{Allard,Bixler,Ferguson}}.  However, these
percentages depend greatly on the corrections made to account for late
type companions that were undetectable by the survey (i.e., M-dwarfs)
and the assumed luminosity of the subdwarfs or observed companions
(i.e., main sequence or subgiant).  Based on their observations of
composite spectrum sdB stars from the PG survey {\citep{PG}},
{\citet{Ferguson}} extrapolated that about 50{\%} of the sdB stars in
the PG catalog had main sequence companions of G8 or later (and a few
had companions earlier than G8).  {\citet{Allard}} identified 31
composite candidates out of 100 sdB stars observed photometrically,
and deduced a fraction of sdB stars with cool main sequence companions
of 54--66{\%}, when allowance was made for perceived selection
effects.  From intermediate resolution spectroscopy of the {\CaII} IR
triplet region of 40 sdB stars, and an assumed ``typical'' luminosity
of a sdB star, {\citet{JeffPoll}} claim that the seven composite
systems they observed contained subgiant companions.  Therefore they
conclude that the the binary fraction must be much higher than what
{\citet{Allard}} concluded based on assuming main sequence companions.
From a sample of 40 UV selected hot subdwarfs, {\citet{Bixler}} found
that eight showed composite spectra, and calculated that 65--100{\%}
of hot subdwarfs had a late type main sequence companion.  However,
our volume limited sample shows significantly fewer composites then
our magnitude limited sample.  Thus, by studying only magnitude
limited samples (or even less well-defined samples) of hot subdwarfs,
it is conceivable that the total number of composite systems could
have been significantly over-estimated by these previous surveys (all
other factors being equal).

The complete 2MASS catalog was released as this paper was nearing
completion.  We will address the full sky sample in a future
contribution.


\acknowledgments

The authors thank E.~M.~Green for providing a list of
spectroscopically observed hot subdwarfs, G.~B.~Berriman for his help
interpreting the 2MASS catalog, and R.~F.~Green for helpful
correspondence.  RW thanks STScI for its hospitality.

This research has been supported in part by: NASA Grid Stars grant
NAG5-9586, NASA GSRP grant NGT5-50399, and a NASA Space Grant
Fellowship through the Pennsylvania Space Grant Consortium.

This publication makes use of: data products from the Two Micron All
Sky Survey, which is a joint project of the University of
Massachusetts and the Infrared Processing and Analysis
Center/California Institute of Technology, funded by the National
Aeronautics and Space Administration and the National Science
Foundation; the USNOFS Image and Catalogue Archive operated by the
United States Naval Observatory, Flagstaff Station
(http://www.nofs.navy.mil/data/fchpix/); the SIMBAD database, operated
at CDS, Strasbourg, France; and the Digitized Sky Surveys.  The
Digitized Sky Surveys were produced at the Space Telescope Science
Institute under U.S.\ Government grant NAG W-2166. The images of these
surveys are based on photographic data obtained using the Oschin
Schmidt Telescope on Palomar Mountain and the UK Schmidt Telescope.
The plates were processed into the present compressed digital form
with the permission of these institutions.


\clearpage


\clearpage

\begin{deluxetable}{rllllcl} 
\tabletypesize{\tiny}
\tablewidth{0in}
\tablecaption{Updated Coordinates and Notes for All Subdwarfs with a
 2MASS 2IDR Identification (ordered by RA). \label{allcoord}}
\tablehead{
  \colhead{No.\tablenotemark{1}} &
  \colhead{KHD\nocite{Kilkenny} ID\tablenotemark{2}} &
  \colhead{RA (J2000)} &
  \colhead{Dec (J2000)} & 
  \colhead{Type} &
  \colhead{Notes\tablenotemark{3}} &
  \colhead{SIMBAD ID\tablenotemark{4}}
 }
\startdata
1525 & CSHK~22957$-23$ & 00 01 32.285 & $-05$ 19 16.76 & sdB    & \ldots                 & BPS~CS~22957$-23$ \\
1526 & CSHK~22876$-28$ & 00 01 37.625 & $-35$ 39 53.09 & sdB    & \ldots                 & BPS~CS~22876$-28$ \\
1527 & PG~23595$+1942$ & 00 02 08.520 & $+19$ 59 13.04 & sd     & \ldots                 & PG~2359$+197$ \\
   1 & SB~7            & 00 03 24.371 & $-16$ 21 06.34 & sdB    & \#3 (CSHK~29517$-41$)  & (BPS~CS~29517$-41$) \\
   6 & TON~S~137       & 00 04 30.992 & $-24$ 26 21.15 & sdO    & \#5 (CSHK~29503$-9$)   & (BPS~CS~29503$-9$) \\ 
   8 & CSHK~22876$-35$ & 00 05 56.689 & $-34$ 53 06.49 & sdB    & \ldots                 & BPS~CS~22876$-35$ \\
  10 & TON~S~140       & 00 06 46.257 & $-27$ 20 53.60 & sdOB   & \#11 (CSHK~29503$-2$)  & (BPS~CS~29503$-2$) \\ 
  12 & PHL~678         & 00 07 33.785 & $+13$ 35 58.70 & sdB    & \ldots                 & \ldots \\ 
  19 & TON~S~144       & 00 10 06.925 & $-26$ 12 56.94 & sdO    & \#17 (CSHK~29503$-25$) & (BPS~CS~29503$-25$) \\ 
  22 & PHL~2726        & 00 12 27.751 & $+03$ 54 31.64 & sd (B) & \ldots                 & \ldots \\ 
\enddata
\tablecomments{
  The complete version of this table is in the electronic edition of
  the Journal.  The printed edition contains only a sample.
 }
\tablenotetext{1}{\tiny
  The sequence number, assigned by us, is not intended to be used as a
  new designator in the literature, but is useful in navigating the
  KHD\nocite{Kilkenny} catalog and cross-referencing this table with
  Table~\ref{all}.
 }
\tablenotetext{2}{\tiny
  The first identifier that appears in KHD\nocite{Kilkenny} for a given entry.
 }
\tablenotetext{3}{\tiny
   {\#}NNNN (Identifier) = indicates a duplicate entry sequence number
     and the first identifier listed for the duplicate in
     KHD\nocite{Kilkenny} (the duplicate entry is not otherwise listed in
     this table);
   $a$ = {\#}87, {\citet{HSTimaging}} report that this pair is resolved into
     three components with HST: the sdO is the NW star of the ``pair,''
     while the SE ``companion'' star resolves into two components
     separated from the sdO by $3\farcs\!37$ and $4\farcs\!48$.
   $b$ = {\#}284, visual double, companion is M-type (KHD\nocite{Kilkenny});
   $c$ = {\#}293, {\citet{UZHM}} report two close companions (``B'' located
     $\leq 2\farcs\!4$ E, and ``C'' located $\sim 3\farcs\!5$ NW); 
   $d$ = {\#}1337 and 1322 were listed with the same name
     (CSHK~29495$-21$) by KHD\nocite{Kilkenny} when in fact they are two
     different objects: {\#}1322 is named correctly, while {\#}1337 is
     not --- {\#}1337 should read CSHK~29495$-63$ in KHD\nocite{Kilkenny}.
 }
\tablenotetext{4}{\tiny
  The identifiers listed in the last column are consistent with SIMBAD
  nomenclature. For example, the designator CSHK~22876$-35$ used by
  KHD\nocite{Kilkenny} for star {\#}8 in this table is equivalent to
  the SIMBAD designator BPS~CS~22876$-35$, and the designator
  PG~23595$+1942$ used by KHD\nocite{Kilkenny} for star {\#}1527 in
  this table is equivalent to the SIMBAD designator PG~2359$+197$.  If
  no other ID is given in the last column, the identifier from
  KHD\nocite{Kilkenny} (given in column 2) is consistent with SIMBAD.
  In this way, one SIMBAD-compliant designator is available for
  cross-reference.  The exceptions are the ``HS'' star {\#}390
  HS~1000$+4704$, the ``SC'' star {\#}1165 SC~1721$-3338$ from
  {\citet{Reid}}, and the ``LBQS'' stars from {\citet[][{\#}41, 565,
  and 1405]{Berg}} which are not presently recognized by SIMBAD;
  although SIMBAD {\it does} use the LBQS designator for quasars from
  {\citet{HFC}}.  In the case of a duplicate entry, if the duplicate's
  identifier listed in column 6 is not consistent with SIMBAD
  notation, then a SIMBAD-compliant name for the duplicate is given
  enclosed in parentheses.
  }
\end{deluxetable}

\clearpage

\begin{deluxetable}{llrcrcrcrclclp{2cm}}
\rotate
\tabletypesize{\tiny}
\tablewidth{0in}
\tablecaption{Magnitudes and Colors for All Subdwarfs (ordered by
  J2000 RA). \label{all}}
\tablehead{
  \colhead{No.\tablenotemark{1}} & \colhead{ID\tablenotemark{2}} &
  \colhead{$J$} & \colhead{$\sigma(J)$\tablenotemark{3}} &
  \colhead{$H$} & \colhead{$\sigma(H)$\tablenotemark{3}} & \colhead{\Ks} &
  \colhead{$\sigma({\Ks})$\tablenotemark{3}} &
  \colhead{{\JKs}\tablenotemark{3}} &
  \colhead{$\sigma({\JKs})$\tablenotemark{3}} &
  \colhead{$V$ (or $y$)\tablenotemark{4}} & \colhead{\bv} &
  \colhead{Type} & \colhead{Notes\tablenotemark{5}}
 }
\startdata
1525 & CSHK22957$-23$ & 12.684 & 0.023 & 12.305 & 0.035 & 12.209 & 0.030  & $+0.475$    & 0.038  & 13.7B    & $\ldots$ & sdB    & \\ 
1526 & CSHK22876$-28$ & 15.038 & 0.044 & 15.063 & 0.099 & 15.169 & 0.199  & $-0.131$    & 0.204  & 14.3B    & $\ldots$ & sdB    & \\ 
1527 & PG23595$+1942$ & 16.238 & 0.097 & 16.246 & 0.230 & 17.103 & \ldots & $<$$-0.865$ & \ldots & 15.44P   & $\ldots$ & sd     & \\ 
   1 & SB~7           & 12.686 & 0.032 & 12.631 & 0.036 & 12.579 & 0.034  & $+0.107$    & 0.047  & 12.67$y$ & $-0.161$ & sdB    & $c$ \\ 
   6 & TONS~137       & 14.534 & 0.037 & 14.714 & 0.064 & 14.765 & 0.113  & $-0.231$    & 0.119  & 13.87$y$ & $-0.273$ & sdO    & \\
   8 & CSHK22876$-35$ & 14.790 & 0.035 & 14.534 & 0.060 & 14.361 & 0.085  & $+0.429$    & 0.092  & 15.4B    & $\ldots$ & sdB    & \\ 
  10 & TONS~140       & 14.543 & 0.031 & 14.687 & 0.065 & 14.863 & 0.101  & $-0.320$    & 0.106  & 13.97$y$ & $-0.253$ & sdOB   & \\ 
  12 & PHL~678        & 13.301 & 0.042 & 13.313 & 0.040 & 13.343 & 0.048  & $-0.042$    & 0.064  & 13.032   & $-0.076$ & sdB    & $s$, AEA \\
  19 & TONS~144       & 12.782 & 0.030 & 12.568 & 0.024 & 12.557 & 0.030  & $+0.225$    & 0.042  & 12.8$y$  & $-0.15 $ & sdO    & \\ 
  22 & PHL~2726       & 13.329 & 0.030 & 13.397 & 0.030 & 13.426 & 0.046  & $-0.097$    & 0.055  & 13.16$y$ & $-0.125$ & sd (B) & $s$ \\
\enddata
\tablecomments{
  The complete version of this table is in the electronic edition of
  the Journal.  The printed edition contains only a sample. 
 }
\tablenotetext{1}{\tiny
  The sequence number, assigned by us, is not intended to be used as a
  new designator in the literature, but is useful in navigating the
  KHD\nocite{Kilkenny} catalog and cross-referencing this table with
  Table~\ref{allcoord}. 
 }
\tablenotetext{2}{\tiny
  Identifiers listed as they appear in KHD\nocite{Kilkenny}.
  SIMBAD-compliant names are given in Table~\ref{allcoord}.
 }
\tablenotetext{3}{\tiny
    Reported errors are 1$\sigma$.  If no error is listed, then the
    star was not detected in that band, and the quoted magnitude is a
    95{\%} confidence upper limit, for {\JKs} the reported value is
    the maximum color index when an upper limit on {\Ks} is reported,
    and it is the minimum color index for an upper limit on $J$.
 }
\tablenotetext{4}{\tiny
   $y$ = Str\"{o}mgren $y$ magnitude listed
     instead of Johnson $V$;
   P, J, or B = Photographic magnitude listed for $V$;
   colon (:) = uncertain magnitude.
 }
\tablenotetext{5}{\tiny
  $s$ = reported single sdB or sdO star used to determine the
    single hot subdwarf locus (see {\S\ref{sec:ColorColor}} and
    Figure~\ref{Known});
  $c$ = reported composite sdB or sdO star used to determine
    the composite hot subdwarf locus (see {\S\ref{sec:ColorColor}} and
    Figure~\ref{Known});
  {\it CC} = 2MASS magnitudes listed are for a nearby companion only
    (the hot subdwarf was not detected);
  $n$ = excluded from the sample because of a ``non-subdwarf''
    classification;
  AEA = $V$ and {\bv} from {\citet{Allard}};
  BBL = {\#}1076 {\&} 1095 had synthetic measurements of $b$ and $y$
    only, see {\S\ref{sec:visual}};
  LAN = $V$ and {\bv} from {\citet{Landolt}};
  MEA = {\#}320 {\&} 1327, Str\"{o}mgren {\uvby} from~\citet{Mooney};
  TYC = $V$ and {\bv} for {\#}234 are converted from Tycho magnitudes 
    (see {\S\ref{sec:visual}}):  $V_{T}=10.194$, 
    $(B_{T}\!-\!V_{T})=+0.111$ {\citep{Tycho}};
  WMG = $V$ and {\bv} from {\citet{WMG}};
  $a$ = {\#}87, {\citet{HSTimaging}} report this pair is resolved into
   three components with HST: the sdO is the NW star of the ``pair,''
   while the SE ``companion'' star resolves into two components
   separated from the sdO by $3\farcs\!37$ and $4\farcs\!48$;
  $b$ = {\#}203, 2MASS noted a horizontal stripe due to bright star
   affecting the {\it H} magnitude only, however, no stripe or nearby
   bright star could be seen in the 2MASS image;
  $d$ = {\#}284, visual double, companion is M-type (KHD\nocite{Kilkenny})
    (only the companion was detected in the 2MASS 2IDR, thus it was excluded
    from the analysis);
  $e$ = {\#}293, {\citet{UZHM}} report two close companions (``B'' located
    $\leq 2\farcs\!4$ E, and ``C'' located $\sim 3\farcs\!5$ NW);
  $f$ = {\#}311, Str{\"{o}}mgren photometry is from {\citet{Wesemael}}
    but was not listed in KHD\nocite{Kilkenny} even though observations for
    other stars from {\citet{Wesemael}} were included in
    KHD\nocite{Kilkenny};
  $g$ = {\#}428, IR excess possibly due to a wind {\citep{TUM}}.
 }
\end{deluxetable}

\clearpage

\begin{deluxetable}{lc}
 \tablewidth{0in}
 \tablecolumns{2}

 \tablecaption{Breakdown of the Hot Subdwarf sample discussed in
 {\S\ref{sec:Sample}}. \label{Breakdown}}

\tablehead{\colhead{Sub-Sample\hspace*{3cm}} &
           \colhead{Number of Objects}}
\startdata
  \multicolumn{2}{c}{\raisebox{0ex}[0.4ex][2ex]{1527 Objects in
          KHD\nocite{Kilkenny}}} \\ \tableline
  Unrecovered                                         & 57   \\
  Duplicate                                           & 11   \\
  Input to 2MASS 2IDR                                 & 1459 \\
 \tableline
  \multicolumn{2}{c}{\raisebox{0ex}[3.5ex][2ex]{593 IDs from 1459
          input to 2MASS 2IDR}} \\ \tableline
  Subdwarfs                                           & 571 \\
  Non-subdwarfs                                       &  22 \\
 \tableline
  \multicolumn{2}{c}{\raisebox{0ex}[3.5ex][2ex]{571 Hot Subdwarfs
          from 2MASS 2IDR}} \\ \tableline
  With 2MASS $J$ and {\Ks}                            & 385 \\
  With 2MASS $J$ but 95{\%} upper confidence on {\Ks} & 180 \\
  Detection of close companion to subdwarf only       &   6 \\
   \tableline
  With $V$ (or $y$)                                   & 348 \\
  With {\BV} (or {\uvby})                             & 339 \\
 \tableline
  \multicolumn{2}{c}{\raisebox{0ex}[3.5ex][2ex]{385 with $J$ and
          {\Ks}}} \\ \tableline
  sdB/sd                                              & 294 \\
  sdO/sdOB                                            &  91 \\
   \tableline
  With $V$ (or $y$)                                   & 277 \\
  With {\BV} (or {\uvby})                             & 273 \\
 \tableline
  \multicolumn{2}{c}{\raisebox{0ex}[3.5ex][2ex]{277 with $V$ (or $y$),
          $J$, and {\Ks}}} \\ \tableline
  sdB/sd                                              & 206 \\
  sdO/sdOB                                            &  71 \\
 \tableline
  \multicolumn{2}{c}{\raisebox{0ex}[3.5ex][2ex]{273 with {\BV}
          (or {\uvby}), $J$, and {\Ks}}} \\ \tableline
  sdB/sd                                              & 202 \\
  sdO/sdOB                                            &  71 \\

\enddata
\end{deluxetable}

\clearpage

\begin{deluxetable}{lccccc}
 \tablewidth{0in}
 \tablecolumns{6}

 \tablecaption{Breakdown of hot subdwarf colors based on {\JKs} for
 the entire sample discussed in {\S\ref{sec:JKs}}. \label{Fractions}}

 \tablehead{ \colhead{ } & \multicolumn{3}{c}{Numbers} &
   \multicolumn{2}{c}{Percentages\tablenotemark{1}} \\
   \colhead{Group} & \colhead{Total} & \colhead{Blue\tablenotemark{2}} &
   \colhead{Red\tablenotemark{3}} &
   \colhead{Blue\tablenotemark{2}} & \colhead{Red\tablenotemark{3}}
   }
 \startdata
  sdB    & 294 & 171 & 123 & $58\pm6$\%  & $42\pm6$\%   \\
  sdO    &  91 &  48 &  43 & $53\pm10$\%  & $47\pm10$\% \\ \tableline
  \raisebox{0ex}[3.0ex][1.5ex]{Both} & 385 & 223 & 162 & $58\pm5$\% & $42\pm5$\%
 \enddata

\tablenotetext{1}{\footnotesize 
  Errors reported are 95{\%} confidence sampling errors of a binomial
  distribution.
 }
\tablenotetext{2}{\footnotesize 
  ``Blue'' is defined as $\JKs <+0.05$ for the sdBs and the total
  sample, and $\JKs <-0.05$ for sdOs.
 }
\tablenotetext{3}{\footnotesize
  ``Red'' is defined as $\JKs >+0.05$ for the sdBs and the total
  sample, and $\JKs >-0.05$ for sdOs.
 }
\end{deluxetable}

\clearpage

\begin{deluxetable}{lcc}
 \tablewidth{0in}
 \tablecolumns{3}

 \tablecaption{Parameters for the Gaussian fit discussed in
 {\S\ref{sec:Gauss}} and shown in
 Figure~\ref{GaussAll}. \label{GaussParameters}}

 \tablehead{ \colhead{Parameter} & \colhead{Blue Gaussian\tablenotemark{1}} &
     \colhead{Red Gaussian\tablenotemark{1}}
   }

 \startdata
   Center    & $-0.158^{+0.004}_{-0.005}$& $+0.311^{+0.014}_{-0.004}$ \\
   Amplitude & $35^{+3}_{-2}$            & $15\pm2$  \\
   Width     & $0.091^{+0.009}_{-0.002}$ & $0.158^{+0.008}_{-0.020}$ \\
   \tableline
   \multicolumn{3}{c}{\raisebox{0ex}[3.0ex][1.5ex]{Integral of Best Fit: 278}}
 \enddata

\tablenotetext{1}{\footnotesize 
  Errors reported are 95{\%} marginal confidence intervals.
 }
\end{deluxetable}

\clearpage

\begin{deluxetable}{lccccc}
 \tablewidth{0in}
 \tablecolumns{6}

 \tablecaption{Breakdown of hot subdwarf colors based on {\JKs} for
 the ``volume limited'' sample discussed in
 {\S\ref{sec:VolColor}}. \label{VolFractions}}

 \tablehead{ \colhead{ } & \multicolumn{3}{c}{Numbers} &
   \multicolumn{2}{c}{Percentages\tablenotemark{1}} \\
   \colhead{Group} & \colhead{Total} & \colhead{Blue\tablenotemark{2}} &
   \colhead{Red\tablenotemark{3}} &
   \colhead{Blue\tablenotemark{2}} & \colhead{Red\tablenotemark{3}}
   }
 \startdata
  sdB    & 228 & 170 & 58 & $75\pm6$\%    & $25\pm6$\%    \\
  sdO    &  72 &  45 & 27 & $62.5\pm11$\% & $37.5\pm11$\% \\ \tableline
  \raisebox{0ex}[3.0ex][1.5ex]{Both} & 300 & 219 & 81 & $73\pm5$\% & $27\pm5$\%
 \enddata

\tablenotetext{1}{\footnotesize 
  Errors reported are 95{\%} confidence sampling errors of a binomial
  distribution.
 }
\tablenotetext{2}{\footnotesize 
  ``Blue'' is defined as $\JKs <+0.05$ for the sdBs and the total
  sample, and $\JKs <-0.05$ for sdOs.
 }
\tablenotetext{3}{\footnotesize 
  ``Red'' is defined as $\JKs >+0.05$ for the sdBs and the total
  sample, and $\JKs >-0.05$ for sdOs.
}
\end{deluxetable}

\clearpage

\begin{deluxetable}{lcc}
 \tablewidth{0in}
 \tablecolumns{3}

 \tablecaption{Parameters for the Gaussian fit to the ``volume
 limited'' sample discussed in {\S\ref{sec:VolColor}} and shown in
 Figure~\ref{Gauss}. \label{VolGaussParameters}}

 \tablehead{ \colhead{Parameter} & \colhead{Blue Gaussian\tablenotemark{1}} &
     \colhead{Red Gaussian\tablenotemark{1}}
   }

 \startdata
   Center    & $-0.154^{+0.013}_{-0.006}$ & $+0.276^{+0.002}_{-0.006}$ \\
   Amplitude & $34\pm1$                   & $10\pm1$  \\
   Width     & $0.096^{+0.005}_{-0.004}$  & $0.101^{+0.002}_{-0.005}$ \\
   \tableline
   \multicolumn{3}{c}{\raisebox{0ex}[3.0ex][1.5ex]{Integral of Best Fit: 214}}
 \enddata

\tablenotetext{1}{\footnotesize
   Errors reported are 95{\%} marginal confidence intervals.
 }
\end{deluxetable}

\end{document}